\documentclass[fleqn,usenatbib]{mnras}
\usepackage{newtxtext,newtxmath}
\usepackage[T1]{fontenc}
\DeclareRobustCommand{\VAN}[3]{#2}
\let\VANthebibliography\thebibliography
\def\thebibliography{\DeclareRobustCommand{\VAN}[3]{##3}\VANthebibliography}

\usepackage{graphicx}	% Including figure files
\usepackage{amsmath}	% Advanced maths commands
\usepackage{multirow}

\title{High angular resolution evidence of dust traps from deep ALMA Band 3 observations of LkCa15}

\author[A. Sierra et al.]{
Anibal Sierra,$^{1,2}$\thanks{E-mail: anibalsierram@gmail.com}
Paola Pinilla,$^{1}$
Laura M. P\'erez,$^{2}$
Myriam Benisty,$^{3}$
Carolina Agurto-Gangas,$^{2}$
\newauthor
Carlos Carrasco-González,$^{4}$
Pietro Curone,$^{2}$
Feng Long$^{5,6}$
\\
$^{1}$Mullard Space Science Laboratory, University College London, Holmbury St Mary, Dorking, Surrey RH5 6NT, UK\\
$^{2}$Departamento de Astronomía, Universidad de Chile, Camino El Observatorio 1515, Las Condes, Santiago, Chile\\
$^{3}$Max-Planck Institute for Astronomy (MPIA), Königstuhl 17, 69117 Heidelberg, Germany\\
$^{4}$Instituto de Radioastronomía y Astrofísica (IRyA), Universidad Nacional Autónoma de México (UNAM), Campus Morelia, Michoacán, Mexico\\
$^{5}$Lunar and Planetary Laboratory, University of Arizona, Tucson, AZ 85721, USA\\
$^{6}$NASA Hubble Fellowship Program Sagan Fellow\\
}

\date{Accepted 2025 March 4. Received 2025 March 3; in original form 2024 December 17
}

\pubyear{\the\year{}}

\begin{document}
\label{firstpage}
\pagerange{\pageref{firstpage}--\pageref{lastpage}}
\maketitle

% Abstract of the paper
\begin{abstract}
Dust traps are the most promising mechanisms to explain the observed substructures in protoplanetary discs. In this work, we present high-angular resolution ($\sim$60 mas, 9.4 au) and high-sensitivity Atacama Large Millimetre/submillimetre Array (ALMA) observations at 3 mm of the transitional disc around LkCa15. The new data, combined with previous high-resolution observations at $\lambda=0.87,1.3$ mm, make LkCa15 an ideal laboratory for testing the dust trapping mechanism.
We found that the width of the three rings decreases linearly with frequency, and the spectral indices show local minima at the locations of the rings, consistent with dust trap models. Multi-wavelength modelling confirms that the dust surface density and maximum grain size peak at 69 and 101\,au, and suggestive peak at 42 au. The estimated total dust mass is between 13-250 M$_{\oplus}$, depending on the chosen opacity. 
The inner disc shows bright and unresolved emission at 3 mm, exhibiting a spectral index of $\alpha_{1.3-3 \rm mm} = 0.3 \pm 0.37$, and $\alpha_{\rm 3mm-3cm}$ ranging from $-0.1$ to $0.0$. These properties are consistent with free-free emission from an ionised jet or disc wind.
Dust evolution models and radiative transfer calculations suggest that a viscosity coefficient of $\alpha = 10^{-3}$, a fragmentation velocity of 10 m s$^{-1}$, and DSHARP opacities provide the best match to the observed properties.
\end{abstract}

% Select between one and six entries from the list of approved keywords.
% Don't make up new ones.

%%%%%%%%%%%%%%%%%%%%%%%%%%%%%%%%%%%%%%%%%%%%%%%%%%

%%%%%%%%%%%%%%%%% BODY OF PAPER %%%%%%%%%%%%%%%%%%

%------------------------------------------------------------------------
\section{Introduction} \label{sec:introduction}
Our understanding of planet formation has significantly advanced over the past decade, mainly thanks to high angular resolution and sensitive observations of protoplanetary disc at millimetre and infrared wavelengths \citep{Andrews_2018, Benisty_2023}, along with state-of-the-art models capable of reproducing and explaining the physics behind the diverse substructures observed in the dust and gas distribution.

One key discovery has been the prediction \citep[e.g.,][]{Birnstiel_2012, Pinilla_2012} and observational detection \citep[e.g.,][]{vanderMarel_2013, Perez_2014, Casassus_2015} of dust traps, which is proposed as an efficient mechanism to halt or slow down the inward migration of dust grains toward the central star, offering a solution to the fast radial drift problem in planet formation theory \citep{Whipple_1972, Weidenschilling_1977}. To date, evidence of dust traps has been observed in several discs at high angular resolution, either through dust continuum millimetre observations \citep[e.g.,][]{Dullemond_2018, Macias_2021, Sierra_2021, Carvalho_2024}, or gas kinematics \citep[e.g.,][]{Rosotti_2020, Izquierdo_2023}. This observational evidence suggests the presence of forming planets \citep{Pinilla_2012}, dead zones \citep[e.g.,][]{Regaly_2012, Flock_2015, Garate_2021, Delage_2023}, or zonal flows \citep[e.g.,][]{Johansen_2009, Uribe_2011, Simon_2014}.

Examples of discs in which dust traps have been proposed to explain their morphology are those with large inner cavities, also known as ``transitional discs'' \citep[e.g.,][]{Calvet_2002, Andrews_2011, vanderMarel_2023}. One notable example is the disc around LkCa15, a K5 star with a mass of 1.2\,M$_{\odot}$, located in the Taurus star-forming region at 157.2 pc \citep{Gaia_2020, Manara_2014}. Its inner dust cavity was resolved over a decade ago \citep{Pietu_2006, Andrews_2011}, a gas surface density drop, which has been shown to be shallow except in the inner 10 au, has been inferred from CO isotopologue observations 
\citep[][see Figure 14 by Leemker.]{vanderMarel_2015, Leemker_2022A}. The constraints on the gas density drop are fundamental to discerning the origin of the ring structure in transitional discs \citep{Bruderer_2013, vanderMarel_2016, Huang_2024}. In particular, the origin of the ring at 69 au in LkCa15 is consistent with a carving Jovian planet \citep{Huang_2024}.

Millimetre observations (angular resolution between 55-150 mas) have revealed the presence of three rings at 42, 69, and 101 au \citep{Isella_2014, Facchini_2020}. Additionally, suggestive evidence of a potential planet at 42 au has been proposed based on dust accumulation at Lagrangian points, as observed in the continuum emission of Bands 6 (1.3 mm) and 7 (0.87 mm) \citep{Long_2022}. The disc around LkCa15 appears to be a strong candidate where dust traps may accumulate dust particles in its rings, potentially triggering the streaming instability \citep[e.g.,][]{Youdin_2005, Stammler_2019, Flock_2021}, as suggested by \cite{Facchini_2020}, and promoting the formation of planetesimals.

Recent modelling of the dust continuum emission at ALMA Bands 7 and 6, and VLA Band Q (7 mm) by \citet{Sierra_2024b} shows that the dust surface density and maximum grain sizes peak around the rings at 69 and 101 au in LkCa15, which is consistent with dust trap models.
However, the 150 mas (23.6 au) angular resolution, limited by the 7 mm observations, made it challenging to resolve the three rings, and analyse the radial profiles with high contrast between rings and gaps. Long-wavelength observations ($\lambda \gtrsim 1.3$ mm), where optically thin emission is detected, are essential for disentangling degeneracies in radiative transfer modelling (Viscardi et al., in prep.) and should not be excluded from the analysis.

In this work, we present deep observations of the dust continuum emission of the disc around LkCa15 at ALMA Band 3 (97.5 GHz, Section \ref{sec:observations}), at an angular resolution of 60 mas (9.4 au) when imaging with traditional techniques, or $35$ mas (5.5 au) when modelling the visibilities. The Band 3 dust continuum observations are compared with the observations at Band 7 and Band 6 in \cite{Long_2022}, allowing us to study the morphology of the dust continuum emission at optically thin wavelengths at high resolution (Section \ref{sec:results}), analyse its azimuthal asymmetries, and look for observational evidence of Lagrangian structures (Section \ref{sec:residuals}). The multi-wavelength data also allows us to infer physical dust properties by fitting the spectral energy distribution (Section \ref{sec:DustProperties}), and put constraints on the amount of dust and non-dust emission in the inner disc of Band 3 (Section \ref{sec:non-dust}).
Dust evolution models tailored for LkCa15 are presented in Section \ref{sec:DustModel}, and the results are discussed in Section \ref{sec:discussion}. Finally, the conclusions are summarized in Section \ref{sec:conclusions}.

%------------------------------------------------------------------------
\section{Observations}\label{sec:observations}

ALMA observed the disc around LkCa15 in Band 3 (Program project: 2022.1.01216.S, PI: Anibal Sierra) from January 2023 to August 2023. The spectral setup was configured to four dedicated dust continuum spectral windows, covering frequencies between 89.54-91.52 GHz, 91.43-93.42 GHz, 101.54-103.52 GHz, 103.50-105.48 GHz, and at a spectral resolution of 31.25 MHz.

\begin{table*}
    \centering
    \caption{Summary of the ALMA Band 3 execution observations.}    
    \begin{tabular}{cccccc}
    \hline
    \hline 
     UIDs & Config & UTC Time & N$_{\rm ant}$ & Baselines & On-source Time\\
     & &  &      & (m-km) & (min)\\     
     (1) & (2) & (3) & (4) & (5) & (6) \\
    \hline
    uid://A002/X1026621/X41c8 & C43-4  &  2023-01-20 03:33:39 & 44 & 15.3-0.8 & 16.63 \\
    uid://A002/X107abb1/X148e0 & C43-7  & 2023-05-27 16:35:59 & 43 & 27.5-3.7 & 26.17 \\
    uid://A002/X107abb1/X13f33 & C43-7  & 2023-05-27 15:04:05 & 42 & 27.5-3.7 & 26.21 \\
    uid://A002/X10b0d7d/X808b  & C43-10 & 2023-08-07 12:58:41 & 48 & 230.2-16.2 & 22.11 \\
    uid://A002/X10b0d7d/X3c87  & C43-10 & 2023-08-06 11:57:31 & 44 & 230.2-16.2 & 41.90 \\
    uid://A002/X10b0d7d/X34bd & C43-10 & 2023-08-06 13:55:42 & 46 & 230.2-16.2 & 41.93 \\
    uid://A002/X10a67b8/X439b & C43-10 & 2023-07-21 14:52:37 & 46 & 230.2-15.2 & 41.90 \\
    \hline
    \end{tabular}\\
    (1) Execution block ID. (2): ALMA configuration. (3) UTC time at the end of the observation. (4) Number of non-flagged antennas. (5) Baseline coverage. (6) Effective on-source time.
    \label{tab:observations}
\end{table*}

The data consist of one short baseline (SB), 2 medium baselines (MB), and 4 long baselines (LB) executions. Table \ref{tab:observations} shows the details of each execution. The data were calibrated using the standard calibration provided by the observatory, where problematic antennas for each execution were flagged.
The calibrated data were self-calibrated using CASA version 6.4.3.27 \citep{McMullin_2007}. We checked the astrometric alignment between different executions by creating low angular resolution images of each execution and fitting a 2D Gaussian to each dust continuum map. However, the inferred offsets were within 5\% of the final beam size. Alternatively, we also verify that the main substructures (ring-like features) observed in each individual execution spatially align.

During each self-calibration iteration, we choose a Briggs robust parameter of 0.5 to image the new dataset and assume that the brightness distribution consists of emission at different angular scales. Therefore, we use the \texttt{multi-scale} algorithm with angular scales corresponding to point sources, 1 beam size, and 2 beam sizes, and a threshold of 2$\sigma$. The signal-to-noise ratio (SNR) is measured after each self-calibration iteration, and we proceed to apply additional iterations if the SNR improves by at least a factor of 5\%. In each iteration, we use the task \texttt{applymode=`calonly'} within the \texttt{applycal} task to calibrate data only and do not apply flags from solutions.

We first self calibrate the SB execution only. We combine spectral windows and scans using \texttt{combine=`spws,scans'} within the \texttt{gaincal} task. After three rounds of phase self calibrations with solution time intervals equals to \texttt{solint = `inf', `360s', `120s'}, the SNR increased 164\%, 10\%, and 8\% with respect to the previous iteration, respectively. No amplitude self calibration was applied because the SNR did not significantly improve.

Before concatenating the MB and LB data, we check for flux-scale alignment and found that the flux of the SB and LB are consistent within the expected uncertainties (less than 1.5\% of difference). However, the flux from the MB disagree by 20\% compared with that from SB and LB data. The SB and LB executions were observed closer in time to the amplitude measurements of the flux calibrator from the ALMA calibrator monitoring \footnote{\url{https://almascience.eso.org/sc/}}. Therefore, the visibilities of the MB datasets are re-scaled to match the SB and LB flux.
Then we concatenate the self calibrated SB to the re-scaled MB executions and apply one phase self-calibration iteration with a solution time interval equal to \texttt{solint = `inf'}. The SNR of the self calibrated data increased by 5\%. 
Finally, we concatenated the LB executions and tried to apply phase and amplitude self-calibration iterations with different solution time intervals, but the SNR did not significantly improve. We create a final Band 3 dust continuum map by cleaning the self calibrated data with a Briggs robust parameter of 0 and a Gaussian uvtaper of 27 mas $\times$ 42 mas; 0 deg, resulting in a beam size of 59 mas $\times$ 55 mas. Then, a convolution was applied using the task \texttt{imsmooth} in \textsc{CASA} to get a circular beam of 60 mas.

In this paper, we compare the Band 3 dust continuum observations of LkCa15 with previous Band 6 (224 GHz) and 7 (340 GHz) observations of the disc. We use the self-calibrated measurement sets from \cite{Long_2022} and re-image the Band 6/Band 7 data using a Briggs robust value of 0.0/0.5 and a uvtaper of 12 mas $\times$ 39 mas; 0 deg/10 mas $\times$ 30 mas; 0 deg. We then apply convolution with the \texttt{imsmooth} task to match the circular 60 mas beam of the Band 3 map. 
The total fluxes in our Band 6 and 7 images are $145.6 \pm 4.3$ mJy and $418.1 \pm 4.7$ mJy, respectively (Section \ref{sec:non-dust}). These values are consistent within the 10\% flux calibration uncertainties when compared with the fluxes of $140 \pm 3$ mJy at 214 GHz reported by \citet{Pietu_2006} using the IRAM PdBI array, $128 \pm 5$ mJy at 224 GHz by \citet{Isella_2012} with CARMA, and $167 \pm 6$ mJy at 230 GHz and $430 \pm 30$ mJy at 340 GHz by \citet{Andrews_2005} using the Sub-millimetre Common-User Bolometer Array (SCUBA).

%--------------------------------------------------------
\section{Results}\label{sec:results}

\subsection{Dust continuum images}\label{sec:multi_observations}
\begin{figure*}
    \centering
    \includegraphics[width=\textwidth]{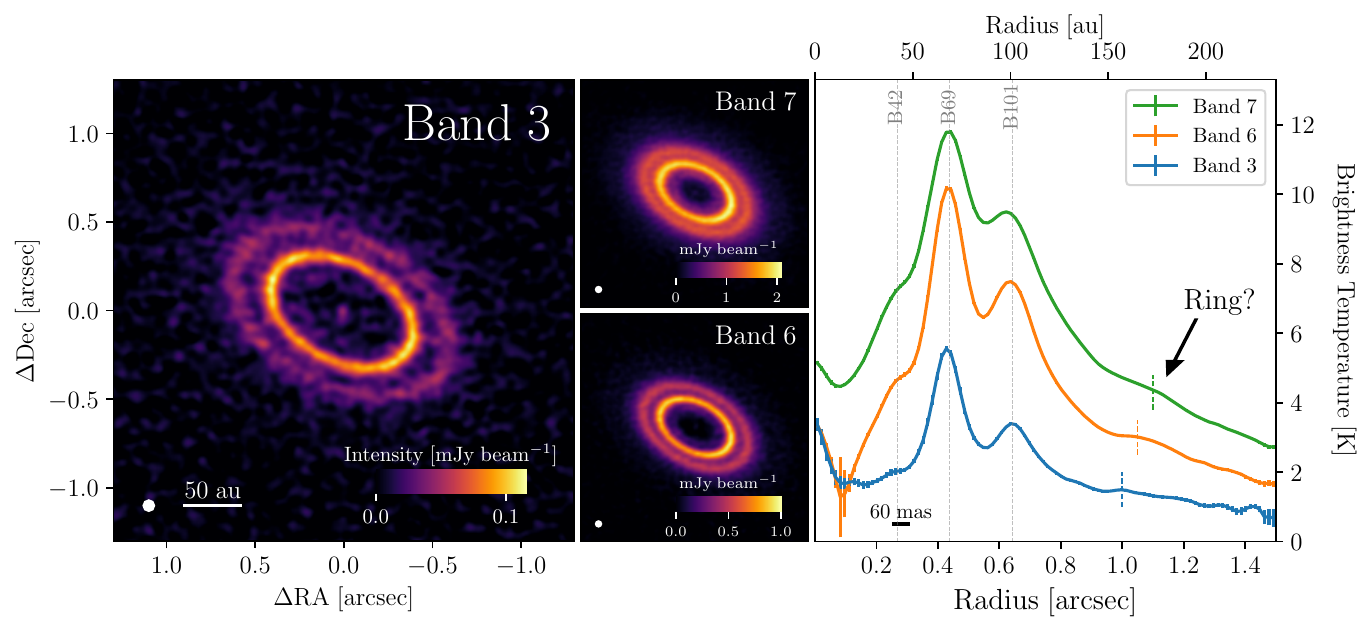}
    \caption{Left: CLEAN image of the dust continuum data at ALMA Band 3. Middle: CLEAN image of the dust continuum data at ALMA Band 7 (Top) and Band 6 (Bottom). Right: Brightness temperature radial profile of the dust continuum data. The circular beam of 60 mas for all maps is shown in the bottom left corner of each panel. The vertical grey dashed lines indicate the positions of the three bright rings (B). The vertical coloured dashed lines indicate the position of a tentative faint ring.}
    \label{fig:CleanData}
\end{figure*}

Figure \ref{fig:CleanData} shows the dust continuum maps in this work. The left panel shows the new Band 3 observations, while the middle panels show the re-imaging data at Band 6 and 7 presented in \cite{Long_2022}. The right panel is the brightness temperature radial profile (without assuming Rayleigh-Jeans approximation) for each map, computed with \textsc{Gofish} \citep{GoFish}. The bright three rings at 42, 69, and 101 au \citep{Long_2022} are distinguishable in the three wavelengths (we follow the nomenclature B42, B69, and B101 for Bright rings). The contrast between the rings and gaps clearly increases with wavelength, suggesting a dust trap origin \citep{Pinilla_2012, Long_2020}. In addition, the Band 3 data shows a bright inner point source, with a size similar to that of the beam.
%, presumably tracing dust + free-free emission.
In the three cases, there is a tentative faint ring in the outer ring, but a different radius: 1.00, 1.05, and 1.10 arcsec at Band 3, 6, and 7, respectively. This faint ring was also identified in the radial profile of the dust continuum observations of LkCa15 at 7 mm (VLA Band Q), located at a radius of 0.99 arcsec \citep{Sierra_2024b}.

The adopted geometry of the disc at all wavelengths is fixed to $\rm inc = 50.1$ deg, and $\rm PA = 61.4$ deg. These values were inferred from two independently methodologies using the Band 3 dataset.  Firstly, they were estimated by minimizing the spread of the de-projected visibilities for circular annuli in the visibility space \citep[e.g.,][]{Isella_2019}. This method constrained a inclination and position angle of $50.1^{+0.4}_{-0.2}$ deg, $61.4^{+0.3}_{-0.4}$ deg, respectively. Secondly, we fit an ellipse in the image plane to a 10$\sigma$ iso-contour, which follows the brightest ring at 69 au. This methodology constrained an inclination and position angle of $49.1^{+0.3}_{-0.2}$ deg, $62.0^{+0.3}_{-0.3}$ deg, respectively. Previous estimates of the disc geometry using the Band 6 and 7 data in \cite{Long_2022} and \cite{Sierra_2024b} have found similar results: 50.2 deg, 61.9 deg and 49.7 deg, 62.1 deg for the inclination and position angle, respectively. However, we chose to use our Band 3 estimations because the higher contrast between rings and gaps allows for better visualization and resolution of the disc geometry, and enables us to trace emission closer to the midplane.

The offset of the disc centre with respect to the phase centre in the Band 3 observations is set to $\rm \Delta RA = 0 \ \rm  mas$, $\rm \Delta Dec = 0 \ \rm  mas$. However, we also infer the disc offset by minimizing the imaginary part of the visibilities \citep[e.g.,][]{Isella_2019}, obtaining $\rm \Delta RA = 15.3 \ \rm mas$ and $\rm \Delta Dec = 12.9 \ \rm mas$. Alternatively, by fitting the iso-contour ellipse to the morphology of the main ring, we obtain $\rm \Delta RA = 15.7 \rm mas$ and $\rm \Delta Dec = -6.9 \ \rm mas$. These offsets are only a factor of 4 and 10 times smaller than the beam size ($\sim$ one pixel size), and do not provide robust evidence of eccentricity. We decided to fix offset to 0 because this particular value minimize the azimuthal residuals after subtracting an axi-symmetric model (Section \ref{sec:residuals} and Appendix \ref{app:Multi_Residuals}), and double-check that this choice does not affect the conclusions about the residuals around the 42 au orbit, where the Lagrangian structures were observed in Band 6 and 7. For the Band 6 and 7 data, we use the offsets reported in \cite{Long_2022}.

Table \ref{tab:imaging} shows a summary of the imaging properties for the three wavelengths. Note that compared with the maps presented in \cite{Long_2022} at an angular resolution of 50 mas, the maps in this work are presented at an angular resolution of 60 mas.

\begin{table*}
    \centering
    \caption{Properties of the millimetre continuum images.}
    \begin{tabular}{cccccccc}
    \hline \hline
    ALMA & $\nu$  & Baselines & robust & uv-tapering & rms Noise & Peak & ALMA \\
    Band & (GHz)  & (m-km)    &        &  (mas $\times$ mas; deg)           & ($\mu$Jy beam$^{-1}$) & SNR & Project Code\\
    \hline
    3  & 97.5 & 15 - 16.2  & 0.0 & 27 $\times$ 42; 0 & 5.4 & 21.3 & 2022.1.01216.S\\
    6  & 224.0 & 15 - 13.9  & 0.0 & 12 $\times$ 39; 0 & 9.5 & 101 & Table 1 in \cite{Long_2022} \\
    7  & 340.0 & 33 - 8.5 & 0.5 & 10 $\times$ 30; 0 & 20.1 & 105 & Table 1 in \cite{Long_2022} \\
    \hline
    \end{tabular}
    \newline
    The robust and uv-tapering for each data set result in a beam size close to 60 mas. A final convolution using the task \texttt{imsmooth} in \textsc{CASA} is used to circularize the beam at 60 mas for all maps.
    \label{tab:imaging}
\end{table*}

\subsection{Frankenstein fit}

In order to exploit the highest angular resolution from the data, we extract the disc visibilities at Band 3 using the self-calibrated measurement set and a modified version of the \textit{export\_uvtable} function in \cite{uvplot}, where the baselines are now normalized using the wavelength of each spectral window and channel. Then we model the de-projected visibilities using the non-parametric visibility fit in the Python package \textsc{Frankenstein} \citep{Jennings_2020}. The Band 6 and 7 visibilities in \cite{Long_2022} are also modelled using the new geometry constraints, but found negligible differences compared with the {\textsc{Frankenstein} fit in that work.

Figure \ref{fig:VisData} shows the results from the \textsc{Frankenstein} fit. The left panel shows the Band 3 data (binned to 3k$\lambda$ and 30k$\lambda$) and the \textsc{\textsc{Frankenstein}} model, which clearly follows the morphology of the data. The inset panel shows that the \textsc{Frankenstein} fit is sensitive to the visibility morphology beyond 1M$\lambda$. The hyper-parameters used in the \textsc{Frankenstein} fits are $\alpha = 1.3$ and $w_{\rm smooth} = 10^{-1}$, which are conservative when inferring substructures from the data \cite{Jennings_2020}. $\alpha$ mimics a SNR threshold when modelling the visibilities, and $w_{\rm smooth}$ controls the smoothness to the power spectrum. Similar to the \textsc{Frankenstein} fit in \cite{Long_2022}, the \textsc{Frankenstein} fit is a good representation of the morphology of the data.

The right panel of Figure \ref{fig:VisData} shows the brightness temperature radial profile computed from the \textsc{Frankenstein} fit. The B42, B69, B101 rings are clearly distinguishable in all cases. In addition, we determined the gap location based on the Band 3 data with the best contrast, which are located at 51, and 83 au (we follow the nomenclature D51, D83 for dark gaps). The brightness temperature in the inner most region of the Band 3 data is higher than that from the Band 6 and 7. The emission from this bright region at Band 3 is consistent with both dust thermal emission and free-free emission, as explored in Section \ref{sec:non-dust}. The tentative faint ring identified from the radial profiles (Figure \ref{fig:CleanData}) is also observed in the radial profiles derived from the \textsc{Frankenstein} fit, with its radial position appearing to decrease with wavelength.

The inset panel zooms in the radial extent around the rings, and shows the normalized intensity for the three analysed wavelengths. It is evident from this plot that the FWHM of the central ring decreases with wavelength, as already suggested from the intensity profile in the image plane (Figure \ref{fig:CleanData}). However, due to differences in the uv-coverage of each observation, we anticipated differences in the intrinsic angular resolution from the \textsc{Frankenstein} fit. 

The angular resolution from the \textsc{Frankenstein} fits is computed following the methodology in \cite{Sierra_2024b}, where the resolution is quantified by fitting the visibilities of an input delta ring at the same uv-coverage of each observation, and then measuring the FWHM of the \textsc{Frankenstein} output profile. The angular resolution from each \textsc{Frankenstein} fit is shown in the bottom of the right panel of Figure \ref{fig:VisData}. The difference in angular resolution is a combination of the longest baselines of each dataset and their SNR. Before comparing the morphology inferred at each wavelength, the radial profiles are convolved using to the lowest angular resolution: 35 mas (5.5 au). 

\begin{figure*}
    \centering
    \includegraphics[width=\textwidth]{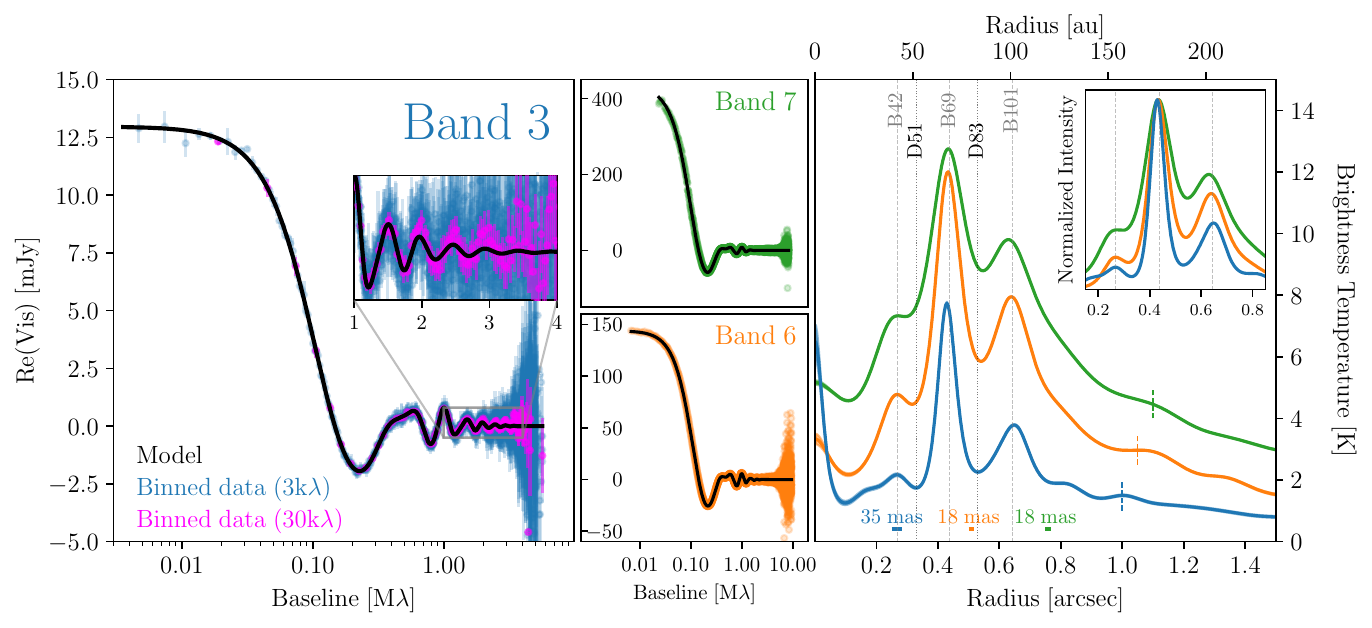}
    \caption{Left: Real part of the continuum visibilities at ALMA Band 3. The plotted data is binned to 3k$\lambda$ (blue) and 30k$\lambda$ (pink) to best visualize the visibility morphology. The \textsc{Frankenstein} model is shown as a solid black line. The inset panel shows a zoom in between 1 and 4 M$\lambda$ in linear scale.
    Middle: Real part of the dust continuum visibilities at Band 7 (Top) and Band 6 (Bottom). The data is shown as coloured dots, and the \textsc{Frankenstein} model is shown as a continuum black line.
    Right: Brightness temperature radial profile computed from \textsc{Frankenstein} fit.  The vertical dashed lines show the position of the three bright rings (B) and dark rings (D). The vertical coloured dashed lines indicate the position of a tentative faint ring.
    The inset panel shows the intensity radial profiles normalized to their peak.}
    \label{fig:VisData}
\end{figure*}

Figure \ref{fig:Widths} shows the normalized intensity radial profiles from the clean image (top row) and from the \textsc{Frankenstein} fit (bottom row) at Band 7, 6 and 3 (from left to right). The angular resolution of the top panels is 60 mas (9.4 au), while the common angular resolution of the bottom panels is 35 mas (5.5 au).

The morphology of the radial profiles in each panel is fitted\footnote{This fit is performed directly on the intensity radial profiles, rather than on the visibilities or the two-dimensional image.} using five symmetric Gaussian rings (grey lines). This model is chosen to account for the three main rings, the inner disc, and the extended emission beyond the outer ring. The free parameters of each Gaussian ring are the intensity amplitude ($A_i$), the width (FWHM$_i$), and the radial location of the peak ($r_i$). The only exception is the Gaussian ring representing the inner disc, where the radial location if fixed at 0. The fit is performed using \textit{curve\_fit}, a non-linear least squares minimisation method based on minimising the chi-squared statistic, implemented in the \text{scipy.optimize} module \citep{SciPy_2020}. Initial guess values were introduced based on the known locations of the rings, their relative flux amplitudes, and an initial FWHM guess of 0.15 arcsec for all rings. The lower boundary for each parameter was set to 0, while no upper boundaries were applied.
The best fit from simple model is able to reproduce the intensity radial profiles for all wavelengths, as shown by the dashed black line of each panel, which is the sum of the individual Gaussian rings.

Table \ref{tab:gaussians} shows the best fit parameters and their uncertainties for the Gaussian functions describing the intensity radial profiles, and the FWHM is plotted in the right panels of Figure \ref{fig:Widths}. The best fit parameters are similar to those reported in Table 2 of \cite{Long_2022}. The FWHM decreases with wavelength for B42, B69, B101, and seem to follow a linear relation. The slopes of the linear functions (fitted using \textit{curve\_fit} without any initial guess values or parameter space boundaries) for the rings B42, B69, B101 in the bottom right panel (radial profiles from the \textsc{Frankenstein} fit), are $0.32 \pm 0.02$ mas GHz$^{-1}$, $0.23 \pm 0.03$ mas GHz$^{-1}$, and $0.42 \pm 0.02$ mas GHz$^{-1}$, respectively. These slopes will be interpreted in the context of dust trapping efficiency in Section \ref{sec:Efficiency}.

\begin{figure*}
    \centering
    \includegraphics[width=\textwidth]{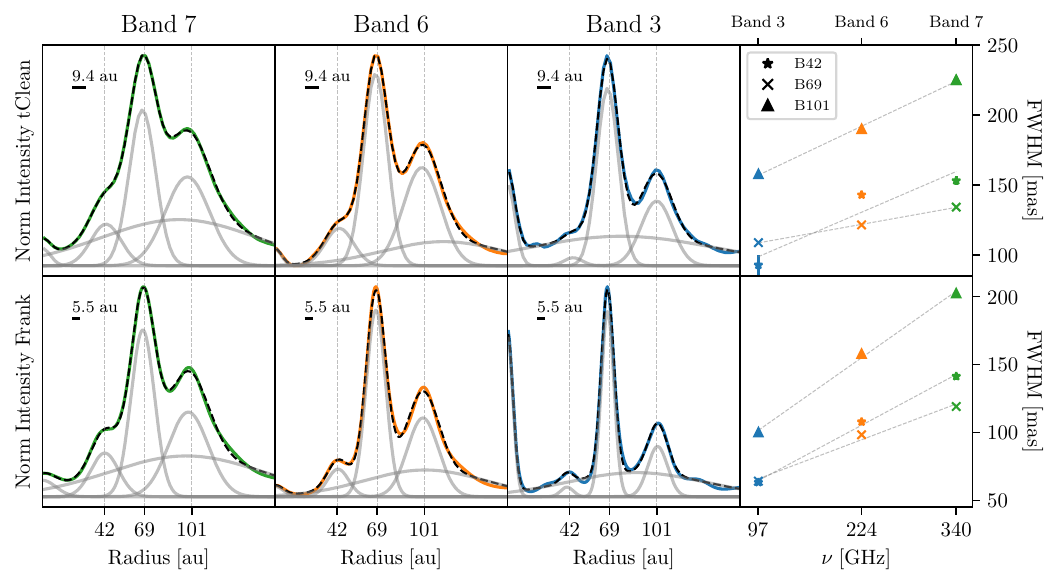}
    \caption{Ring morphology of LkCa15 at ALMA Band 7 (first column), Band 6 (second column), and Band 3 (third column) computed from the azimutally averaged radial profile in the image plane (first row) and \textsc{Frankenstein} fit (second row). 
    The angular resolution of the top and bottom panels is 60 mas (9.4 au) and 35 mas (5.5 au), respectively, and it is shown as a horizontal black line in the top left of each panel.
    The grey lines are the best fit for individual Gaussian profiles (fitted to the intensity radial profiles), and the thin black line shows their sum. Three vertical dashed lines are included as reference at the ring locations of 42, 69, and 101 au.
    The last column shows the Full-Width-Half-Maximum of the three main rings (see marker code in the top right panel) at different wavelengths.}
    \label{fig:Widths}
\end{figure*}

\begin{table*}
    \caption{Best fit parameters of the Gaussian rings reproducing the intensity radial profiles computed from the image plane (Clean columns) and from Frankenstein fit (Frankenstein columns)}.
    \centering
    \begin{tabular}{c|ccc|ccc}
    \hline 
    \multirow{2}{*}{Parameter} & \multicolumn{3}{c|}{Clean} & \multicolumn{3}{c}{Frankenstein} \\
                              & B7               & B6               & B3               & B7               & B6               & B3 \\
    \hline \hline
    $r_{0}^{*}$               & 0                & 0                & 0                & 0                & 0                & 0 \\
    FWHM$_{0}$ [au]          & $12.6 \pm 1.0$   & $10.1 \pm 0.7$   & $11.8 \pm 0.3$   & $21.2 \pm 2.2$   & $13.1 \pm 1.3$   & $8.6 \pm 0.1$ \\
    $\log_{10}(A_{0})$        & $-0.8 \pm -2.1$  & $-1.2 \pm -2.4$  & $-1.5 \pm -3.2$  & $9.2 \pm 1.0$    & $8.7 \pm 1.0$    & $9.0 \pm 1.0$ \\
    \hline
    $r_{42}$ [au]             & $42.9 \pm 0.4$   & $43.5 \pm 0.3$   & $44.4 \pm 1.0$   & $42.1 \pm 0.2$   & $42.6 \pm 0.2$   & $40.2 \pm 0.3$ \\
    FWHM$_{42}$ [au]         & $24.1 \pm 1.1$   & $22.5 \pm 0.9$   & $14.6 \pm 2.8$   & $22.2 \pm 0.9$   & $17.0 \pm 0.6$   & $10.0 \pm 0.8$ \\
    $\log_{10}(A_{42})$       & $-0.4 \pm -1.9$  & $-0.8 \pm -2.2$  & $-2.5 \pm -3.3$  & $9.7 \pm 8.3$    & $9.2 \pm 7.7$    & $7.8 \pm 6.6$ \\
    \hline
    $r_{69}$ [au]             & $67.5 \pm 0.1$   & $68.0 \pm 0.1$   & $67.6 \pm 0.1$   & $67.8 \pm 0.1$   & $68.2 \pm 0.1$   & $67.6 \pm 0.1$ \\
    FWHM$_{69}$ [au]         & $21.1 \pm 0.2$   & $19.1 \pm 0.1$   & $17.1 \pm 0.1$   & $18.7 \pm 0.1$   & $15.5 \pm 0.1$   & $10.1 \pm 0.1$ \\
    $\log_{10}(A_{69})$       & $0.1 \pm -2.0$   & $-0.1 \pm -2.2$  & $-1.1 \pm -3.3$  & $10.2 \pm 8.2$   & $10.0 \pm 7.8$   & $9.1 \pm 6.7$ \\
    \hline
    $r_{101}$ [au]            & $97.9 \pm 0.2$   & $99.3 \pm 0.1$   & $100.8 \pm 0.1$  & $98.6 \pm 0.1$   & $100.0 \pm 0.1$  & $101.5 \pm 0.1$ \\
    FWHM$_{101}$ [au]        & $35.4 \pm 0.5$   & $29.9 \pm 0.4$   & $24.8 \pm 0.5$   & $31.8 \pm 0.4$   & $24.8 \pm 0.3$   & $15.8 \pm 0.2$ \\
    $\log_{10}(A_{101})$      & $-0.1 \pm -2.0$  & $-0.4 \pm -2.2$  & $-1.6 \pm -3.3$  & $9.9 \pm 8.1$    & $9.6 \pm 7.7$    & $8.5 \pm 6.6$ \\
    \hline
    $r_{\rm ext}$ [au]        & $92.5 \pm 1.0$   & $113.9 \pm 3.2$  & $77.1 \pm 1.5$   & $97.3 \pm 2.3$   & $101.5 \pm 1.4$  & $87.0 \pm 0.6$ \\
    FWHM$_{\rm ext}$ [au]    & $124.8 \pm 1.6$  & $97.0 \pm 3.3$   & $150.0 \pm 4.6$  & $122.6 \pm 2.1$  & $96.0 \pm 1.7$   & $109.8 \pm 1.5$ \\
    $\log_{10}(A_{\rm ext})$  & $-0.4 \pm -2.0$  & $-1.0 \pm -2.3$  & $-1.9 \pm -3.3$  & $9.6 \pm 8.1$    & $9.1 \pm 7.7$    & $8.2 \pm 6.4$ \\
    \hline
    \end{tabular}
    \newline 
    $^{(*)}$: fixed value. Amplitude units: [$A_i$ from Clean] = mJy beam$^{-1}$, [$A_i$ from visibilities] = Jy sr$^{-1}$.
    \label{tab:gaussians}
\end{table*}

%--------------------------------------------------------
\subsection{Band 3 azimuthal structures}\label{sec:residuals}
Recently, \cite{Long_2022} reported azimuthal structures in the Band 6 and Band 7 dust continuum emission inside the disc cavity of LkCa15. These azimuthal structures, separated by $\sim 120$ deg, are consistent with the Lagrangian Points of a forming $\sim$Neptune-mass planet located at 42 au from the central star, and may be tracing local dust traps. 

We search for similar azimuthal residuals in the Band 3 dust continuum data by creating a measurement set of the Band 3  \textsc{Frankenstein} model, along with a measurement set for the residuals (observed visibilities minus model visibilities). These datasets are generated from the original measurement sets by replacing the observed visibilities with either the \textsc{Frankenstein} model or the visibility residuals. They are then imaged in \textsc{CASA} using the same parameters as in Table \ref{tab:imaging}.

It is known that the residuals can be strongly affected by the adopted disc geometry parameters \citep{Andrews_2021}. As mentioned in Section \ref{sec:multi_observations}, the offset of the disc centre with respect to the phase centre for the Band 3 data is fixed to $\Delta$RA = $\Delta$Dec = 0. This is because the structures in the residual map are minimized for these values (Appendix \ref{app:Multi_Residuals}). Figure \ref{fig:Robust-Residuals} shows the clean image map of the \textsc{Frankenstein} Model and Residual. The model is a good representation of the axi-symmetric structure of LkCa15 at Band 3, as shown by the low residuals ($< 3$ SNR). However, there are some azimuthal residuals with a SNR above 3 or below -3, specially within the disc cavity, as shown in the zoom-in residual maps in the middle-right panel.

We detected a point source at the 42 au orbit with an SNR of 3.5, which persists independently of the disc offset (Appendix \ref{app:Multi_Residuals}). This point source lies near the border of the Northern arc identified in \cite{Long_2022} (Right panel of Figure \ref{fig:Robust-Residuals}). This point source is not detected at Band 6 or 7, therefore, its upper limit flux at Band 6 and 7 are the rms values of $9.5$ and $20.1 \ \mu$Jy, respectively. For reference, the circumplanetary disc around PDS70c (scaled to the LkCa15 distance) is $44 \pm 8$ $\mu$Jy at Band 7 \citep{Benisty_2021}, only $\sim 2$ the rms of the LkCa15 Band 7 observations. The absence of Lagrangian structures in Band 3 is discussed in Section \ref{sec:NoLagrangian}.

Two planet candidates at orbital distances of 16.5 au and 20.9 au were proposed based on infrared and H$\alpha$ observations (February 7, 2015) reported by \cite{Sallum_2015}. However, these were later explained as scattered light (at the same orbital radii) from the forward-scattering side of the disc, rather than the presence of orbiting companions \citep{Sallum_2023}. The apparent position angle of these structures changes in the opposite direction to the disc rotation.
In our data, we only detect positive residuals with low SNR of 3.0 in the 16.5 au orbit, and SNR of 2.5 in the 20.9 orbit (blue and grey iso-contours in the middle-right and right panels of Figure \ref{fig:Robust-Residuals}). When estimating the Keplerian rotation around the 1.2 M$_{\odot}$ star for each orbit (opposite to the disc rotation), only the outer point source partially matches the expected position. However, given the low SNR, these detected residuals cannot be considered evidence of continuum emission associated with these planet candidates.

\begin{figure*}
    \centering
    \includegraphics[width=\textwidth]{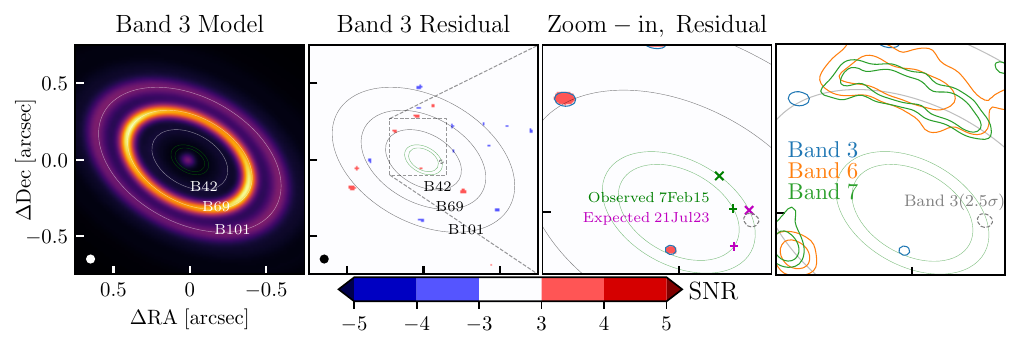}
    \caption{Left: Clean image of the \textsc{Frankenstein} model. The colour bar is the same as the left panel of Figure \ref{fig:CleanData}.
    Middle left: Clean image of the visibility residuals.
    Middle right: Zoom-in to the residual map within the cavity.
    %The iso-contours are at 3$\sigma$ level.
    Right: Iso-contours of the residual maps at Band 3 (blue, at 3$\sigma$ level), Band 6 (orange, at 5$\sigma$, 10$\sigma$, 12$\sigma$ level), and Band 7 (green, at 3$\sigma$, 5$\sigma$ level). The dashed grey iso-contour shows one Band 3 residual at 2.5$\sigma$ level.
    The white and black ellipses show the radial position of the three bright rings at 42, 69, and 101 au, while the green ellipses are reference orbits at 16.5 and 20.9 au.
    The green x and + markers in the zoomed Residual panel indicate the position of the scattered-light and H$\alpha$ detection in February 2015. Meanwhile, the purple markers show their expected positions following a Keplerian orbit (in the opposite direction of the disc rotation; see main text) as of the date of the LB Band 3 observations (July 2023).}
    \label{fig:Robust-Residuals}
\end{figure*}

%--------------------------------------------------------
\subsection{Dust properties}\label{sec:DustProperties}
The multi-wavelength observations of LkCa15 can be used to constrain the radial profiles of the dust temperature ($T_{\rm d}$), dust surface density ($\Sigma_{\rm d}$), and maximum grain size ($a_{\rm max}$), as those inferred in several resolved discs \citep[e.g.,][among others]{Carrasco-Gonzalez_2019, Macias_2021, Sierra_2021, Zhang_2023, Carvalho_2024}. 

The spectral indices between Bands 6 and 7 ($\alpha_{\rm B6-B7}$) and between Bands 3 and 6 ($\alpha_{\rm B3-B6}$), shown in Figure \ref{fig:SPI}, provide an initial insight into the dust properties around the rings and gaps of LkCa15. The spectral index $\alpha_{\rm B6-B7}$ coincides with that computed in \cite{Long_2022}. Both radial profiles show three local minima at the location of the main rings, which might be explained by the emission of large grain sizes. In addition, the low spectral index ($\alpha_{\rm B6-B7} \sim 2$) in B69 and B101 also suggests that emission at the two bands are likely optically thick.
The radial profiles of the spectral indices vary significantly in the region within the B42 ring, while $\alpha_{\rm B6-B7}$ tends to large values (possibly indicating small grain sizes), $\alpha_{\rm B3-B6}$ tends to very low values, close to or below 0.

It is known that dust scattering is able to reproduce spectral indices between $\sim 1.5$ and 2.0 using optically thick emission and high albedo grain sizes \citep{Liu_2019, Zhu_2019, Sierra_2020}. Furthermore, the inclusion of vertical temperature structures, scattering, and high albedo grain sizes can reproduce spectral indices as low as 0.5 using only dust thermal emission \citep{Sierra_2020}. This may help explain the physical conditions within the innermost region of LkCa15. However, other non-dust radiative processes, such as compact free-free emission, could also explain spectral indices below 2. The region within 0.1 arcsec (approximately three times the angular resolution), where the spectral index is below 1.5, is shown as a grey shaded area in Figure \ref{fig:SPI}. We note that the innermost region (within 1–2 beam sizes) may have larger uncertainties due to angular resolution limitations or uncertainties in the \textsc{Frankenstein} fitting process \citep[see][]{Jennings_2021}. Consequently, conclusions drawn for this region should be interpreted with caution.
In Section \ref{sec:non-dust}, we study the integrated dust continuum fluxes and spectral indices in the inner disc of LkCa15, and compare them with those of other transitional discs.

\begin{figure}
    \centering
    \includegraphics[width=\linewidth]{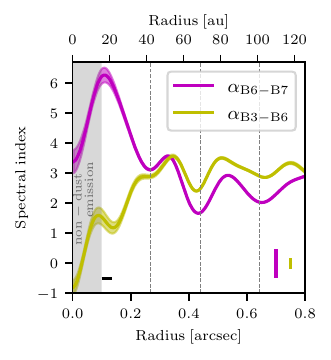}
    \caption{Spectral index radial profile between ALMA Band 6 and 7 (magenta), and ALMA Band 3 and 6 (yellow). 
    The shaded grey within 0.1 arcsec show the region with possible free-free emission at Band 3.
    The vertical dashed lines show the position of the B42, B69, and B101 rings. The beam size (35 mas, 5.5 au) is shown in the bottom left corner as a horizontal black line. The shaded magenta and yellow regions indicate $1\sigma$ uncertainties.
    The vertical bars in the bottom right corner show the error bars from the ALMA absolute flux calibration uncertainty.}
    \label{fig:SPI}
\end{figure}

The dust property constraints are computed using the solution for the radiative transfer equation in \cite{Sierra_2019, Sierra_2020}, where the contribution of the scattering opacity is taken into account. We explore the space parameter between 1 K $< T_{\rm d}< $ 300 K, $10^{-6}$ g cm$^{-2} < \Sigma_{\rm d} < $ 10 g cm$^{-2}$, and 10 $\mu$m $ < a_{\rm max} <$ 30 cm, and use the Markov chain Monte Carlo (MCMC) implemented in the \textsc{Python} library \textsc{emcee} \citep{Foreman_2013}. The $\chi^2$ is defined as in Equation 5 of \cite{Sierra_2024b}, and we also consider the absolute flux uncertainty for each ALMA Band according to the ALMA proposer's guide: 10\% for Band 6 and 7, and 5\% for Band 3.

The prior for the maximum grain size and dust temperature are given by a log-uniform distribution, while for the dust temperature we include a prior temperature radial profile given by the expected temperature from a passively irradiated disc: $T_{\rm prior} = (0.02 L_* /8\pi \sigma_{\rm B} r^2)^{1/4}$, where $\sigma_{\rm B}$ is the Stefan-Boltzmann constant. We follow the Equation 6 in \cite{Sierra_2024b} to describe the prior probability distribution, and use a wide temperature uncertainty around the prior temperature (50 K) to ensure that the prior is not dominating the posterior probability distribution.

The dust opacities are estimated by two models: the DSHARP opacities \citep{Birnstiel_2018}, and Ricci's opacities \citep{Ricci_2010}. In both cases, we consider sphere grains and use the DSHARP opacity tools\footnote{DSHARP opacity tools available at \href{github.com/birnstiel/dsharp_opac}{github.com/birnstiel/dsharp\_opac}}. The DSHARP composition consists on water ice, astronomical silicates, troilite, and refractory organic, with volume fractions of 36.42\%, 16.70\%, 2.58\%, 44.30\% respectively. Meanwhile, the Ricci's opacity composition consists on water ice, astronomical silicates, carbon grains, and vacuum, with volume fractions of 42\%, 7\%, 21\%, and 30\%, respectively. A full comparison between these dust opacity models is discussed in \cite{Birnstiel_2018}. We assume the particle size distribution follows $n(a)da = a^{-p}da$, with $p=3$. This power law lies between the expected values when the maximum grain size is dominated by fragmentation ($p\sim 3.5$) and drift ($p\sim 2.5$), as shown by the dust evolution models in \cite{Birnstiel_2012}.

Finally, the parameter space at each radius is explored using 24 walkers and 10,000 steps. The best fits at each radius are then used to sample the radial profile of dust properties, taking the median value at every 1/3 of a resolution element\footnote{A public version of the code is available at \href{https://github.com/anibalsierram/DustProperties}{github.com/anibalsierram/DustProperties}}. 

The results are shown in Figure \ref{fig:DustProp} for the radial extent where no free-free contamination is expected ($> 0.1$ arcsec). Local maxima for the dust surface density and maximum grain size are inferred at the ring locations in both models; however, their absolute values differ significantly. For DSHARP opacities, grain sizes on the order of millimetres are inferred in the gaps, and centimetre-sized grain sizes in the rings. In contrast, using Ricci's opacities, grain sizes below 1 mm are inferred in both the rings and gaps, with oscillations of tenths of a millimetre between them. This grain size difference has an important impact on the order of magnitude of the dust surface density and total dust mass. While the dust mass inferred from the DSHARP opacities is around $7.5\times 10^{-4}$ M$_{\odot}$ (250 M$_{\oplus}$), for Ricci's opacities it is an order of magnitude lower: $0.4 \times 10^{-4}$ M$_{\odot}$ (13 M$_{\oplus}$). Additional observational constraints from the properties inferred in both opacity models are discussed in Section \ref{sec:opacities}.

The optical depths for both models are estimated and presented in Figure \ref{fig:OptDepths}. We remark that the multi-wavelength analysis includes both absorption and scattering opacities. The contribution of the absorption-only optical depth is shown as a blue curve, while the contribution from both absorption and scattering is shown as an orange curve. The latter is higher for the DSHARP opacity model because the albedo of mm-cm grain sizes is higher than that of sub-mm grain sizes at millimetre wavelengths. 
For both models, the absorption-only optical depth is below 1 everywhere (it is only close to 1 in the brightest ring at Band 7 for the DSHARP opacity model). This indicates that the inclusion of scattering opacities has a small effect on the radiative transfer equation \citep{Sierra_2020}, and the results should be similar if the formal solution to the radiative transfer equation with no scattering is used. The optical depth at Band 6 from both the DSHARP and Ricci's opacity model is similar to that in \cite{Facchini_2020}, where no-scattering effects and the temperature of a passively irradiated disc are assumed.

\begin{figure*}
    \centering
    \includegraphics[width=\textwidth]{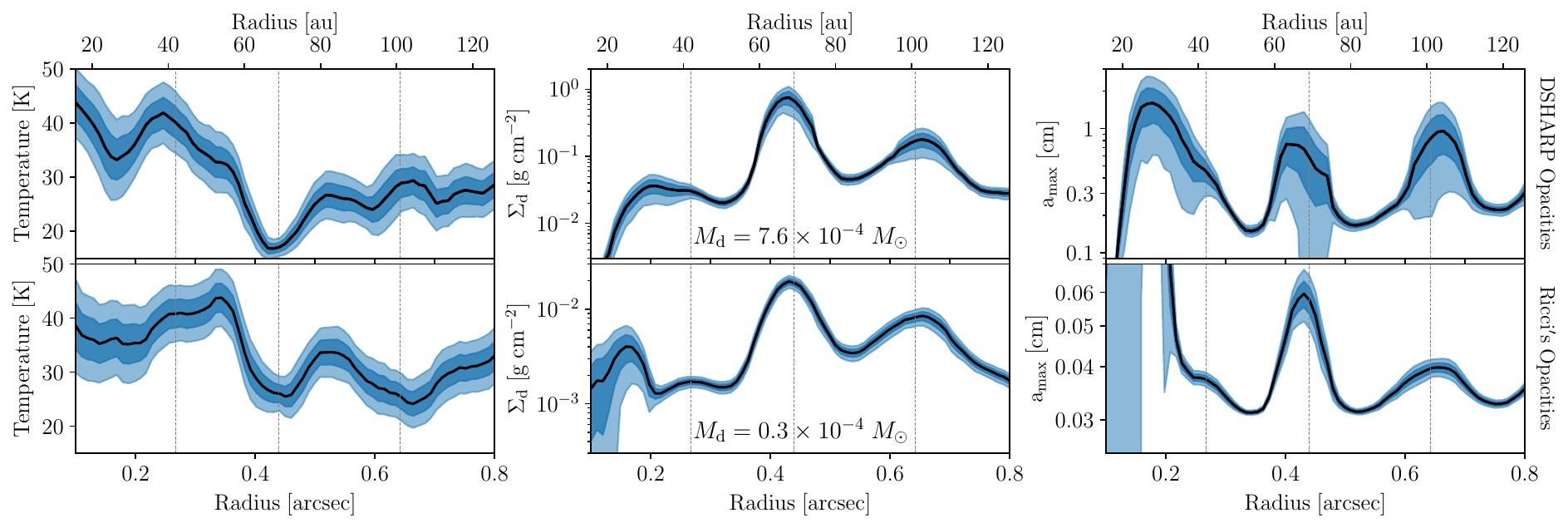}
    \caption{Dust properties constraints from the multi-wavelength dust continuum observations of LkCa15 using the DSHARP opacities (Top panels) and Ricci's opacities (Bottom panels). The shaded regions show 1$\sigma$ and 2$\sigma$ uncertainties. The vertical dashed lines show the position of the ring locations at 42, 69, and 101 au.}
    \label{fig:DustProp}
\end{figure*}

\begin{figure*}
    \centering
    \includegraphics[width=0.85\textwidth]{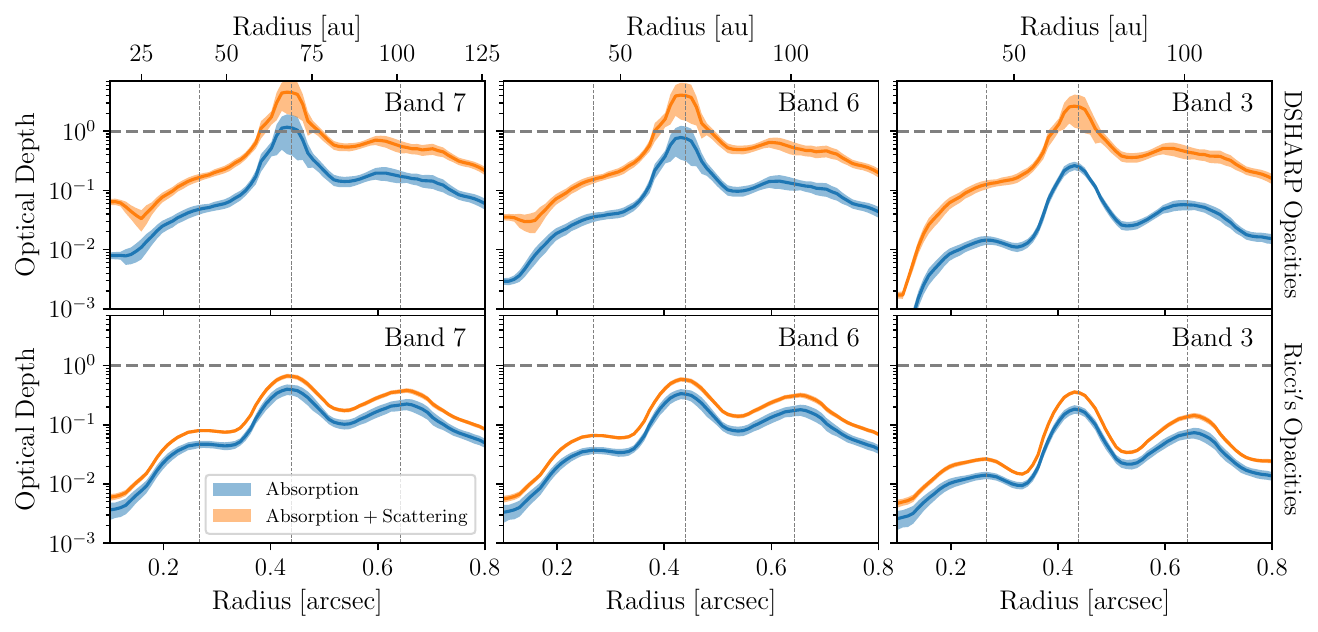}
    \caption{Optical depths at ALMA Band 7, 6 and 3 (from left to right) of LkCa15 using the DSHARP opacities (Top panels) and Ricci's opacities (Bottom panels). The blue line is the absorption only optical depth, while the orange line is the absorption + scattering optical depth. The shaded region shows 2$\sigma$ uncertainties.
    The vertical dashed lines show the position of the ring locations at 42, 69, and 101 au.}
    \label{fig:OptDepths}
\end{figure*}

%----------------------------------------------------------
\subsection{Non-dust emission}\label{sec:non-dust}
As discussed in the previous Section, free-free emission is expected in the inner region ($<0.1$ arcsec) of LkCa15, where the spectral index between Band 6 and 3 is below $\sim 1.5$ \citep[the minimum spectral index that can be explained by dust thermal emission and scattering,][]{Sierra_2020}. 
Previous studies of the compact emission in the centre of the disc around LkCa15 at 7 mm suggest that this emission is not consistent with the radiation from the stellar photosphere, but from ionized gas \citep{Isella_2014}. This was confirmed by the analysis of \cite{Zapata_2017}, where the SED at millimetre and centimetre wavelengths is described by a two-power law fit, accounting for two emission processes: dust continuum emission and free-free emission.

We study the SED of LkCa15 using the integrated fluxes reported in \cite{Zapata_2017}, and including the resolved flux constraints in this work at ALMA Band 7, 6 and 3. The previous flux estimations at ALMA Band 7 and 6 in \cite{Andrews_2005} were updated using our data, but they are consistent within the error bars. Additionally, we include the flux constraints from the ALMA Band 9 and VLA Band Q observations presented in \cite{Sierra_2024b}. In a few cases (ALMA Band 3, 6, and 7, and VLA Band Q), we also analyse the fluxes in the inner disc, where most of the free-free emission is expected. 

The total fluxes were computed using the flux curve of growth, extracting the asymptotic flux at large radii where the integrated flux oscillates only around the RMS value. In all cases, we use the intensity radial profiles computed from the image plane to create the flux curve of growth.  We did not use the intensity profiles derived from the \textsc{Frankenstein} fits, as constraints on the inner region can have substantial uncertainties \citep{Jennings_2020, Jennings_2021}. 
The fluxes in the inner disc are taken from \citet{Isella_2014} and \citet{Long_2022}, while the Band 3 flux is estimated from the peak value in the compact inner disc region, which is consistent with the result obtained using the \texttt{imfit} task in \textsc{CASA}.
Table \ref{tab:Fluxes} shows a summary of the fluxes at different bands, and Figure \ref{fig:SED} shows the SED computed from the total fluxes and the inner disc.

\begin{table}
    \centering 
    \caption{Flux densities at millimetre and centimetre wavelengths of LkCa15.}
    \begin{tabular}{cccc}
    \hline
    Frequency  & ALMA/VLA  & Total Flux & Inner disc \\
    $\rm [GHz]$ & Band & [mJy]  & [$\mu$Jy] \\
    \hline \hline
    689 & 9 & 1436.9 $\pm$ 72.4 & - \\
    341 & 7 & 418.1 $\pm$ 4.7 & 155.8 $\pm$ 7.9 \\
    228 & 6 & 145.6 $\pm$ 4.3 & 58.6 $\pm$ 4.1  \\ 
    107 & 3 & 17.0 $\pm$ 0.8 & - \\    
    97.5 & 3 & 13.2 $\pm$ 1.9 & 45.4 $\pm$ 14.0\\
    44.0 & Q & 0.82 $\pm$ 0.18 & 16.6 $\pm$ 3.6 \\      
    9.10 & X & 0.045 $\pm$ 0.017 & - \\
    5.00 & C & $\leq 2.1$E-2 & - \\
    \hline
    \end{tabular} \\ 
    References: \cite{Zapata_2017} for Band X and C. \cite{Pietu_2006} for Band 3 at 107 GHz. \cite{Long_2020} for the inner disc at Band 7 and 6, and \cite{Isella_2014} for Band Q.
    All other fluxes were computed in this work.
    \label{tab:Fluxes}
\end{table}

\begin{figure}
    \centering
    \includegraphics[width=0.5\textwidth]{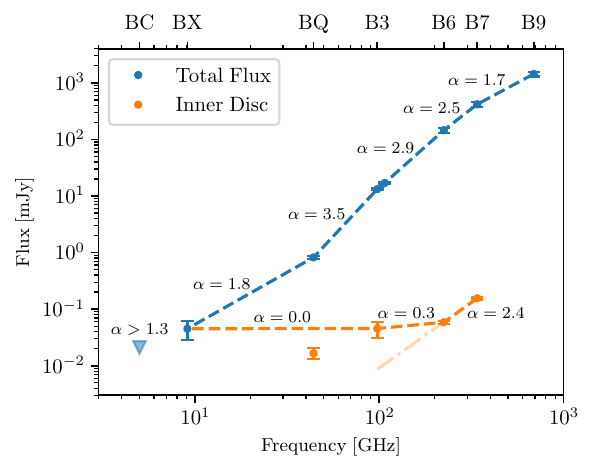}
    \caption{Spectral Energy Distribution (SED) of LkCa15 at millimetre and centimetre wavelengths computed from the total flux (blue), and inner disc (orange). The triangle at Band C represents the upper limit detection at that frequency. The faint orange line between Bands 6 and 3 in the inner disc represents the extrapolated flux using the spectral index between Bands 6 and 7.
    The spectral indices between consecutive bands are indicated in each case. }
    \label{fig:SED}
\end{figure}

The total fluxes clearly follow different power laws below and above Band Q. The spectral indices increase with wavelength, from 1.7 between Band 9 and Band 7 to 3.5 between Band 3 and Band Q, as expected for dust continuum emission. However, the spectral index decreases to 1.8 between Band X and Band Q, suggesting contributions from sources other than dust thermal emission, such as free-free or synchrotron emission. The upper-limit detection at Band C provides a lower limit for the spectral index of 1.3 between Band C and Band X, suggesting that the free-free emission could be partially optically thick between such wavelengths.

On the other hand, the spectral indices in the inner disc appear to follow a different behaviour. The spectral index between Bands 6 and 7 is $2.4 \pm 0.3$ \citep{Long_2022}, decreasing to $\alpha = 0.30 \pm 0.37$ between Bands 3 and 6. The latter is similar to what is expected from free-free emission, which typically ranges from $-0.1$ (at high frequencies) to $2.0$ (at low frequencies), depending on the optical depth regime \citep{Rybicki_1986}.

Recently, \cite{Rota_2024} studied the spectral index between millimetre and centimetre wavelengths in the inner discs of 11 transitional discs, demonstrating that this value is lower ($<$2) compared to the spectral indices in the outer discs ($>$2). In several discs, the low spectral indices in the inner disc extend up to Band 6, as also observed in the case of LkCa15. 
\cite{Rota_2024} suggest that the difference in spectral indices between the inner and outer discs arises because the emission in the former includes a significant contribution from free-free emission associated with ionised jets or disc winds.

We follow the methodology of \cite{Rota_2024} to constrain the upper and lower limits of the spectral index between Band 3 and the centimetre observations in Band X in the inner disc. Assuming that the emission in Bands 6 and 7 originates from dust continuum emission, the extrapolated dust continuum flux in Band 3 is approximately $8.6 \mu$Jy, which represents $\sim 19$\% of the inner disc flux in Band 3. Consequently, the estimated free-free emission in the inner disc at Band 3 is $\sim 36.8 \mu$Jy, and the lower limit for the spectral index between Band 3 and Band X (where the total flux is presumably dominated by the inner disc) is $\alpha_{\rm extrap} = -0.1$. The upper limit is calculated by determining the spectral index without subtracting the dust continuum contribution in Band 3, yielding an upper limit of $\alpha_{\rm 3mm-3cm} = 0.0$

These spectral indices constraints are similar as those found for other transitional discs analysed in \cite{Rota_2024}. In particular, they are similar to the constraints in GG Tau, HD 100546, and HD 142527, where the emission from the inner disc is also detected in both Bands 6 and 7. \cite{Rota_2024} also found that the ionised mass loss rate and accretion rate onto the central star of LkCa15 follow the same relation as those of the other discs in their sample, establishing LkCa15 as another source where the origin of free-free emission can be attributed to an ionised jet. Suggestive evidence of unresolved free-free emission in the inner disc at Band 3 has also been recently reported for the disc around PDS 70 \citep{Doi_2024}, where the spectral index between Bands 7 and 3 is low ($\alpha < 2$).

%----------------------------------------------------------
\section{Dust evolution models for LkCa15} \label{sec:DustModel}

\begin{figure*}
    \centering
    \includegraphics[width=\textwidth]{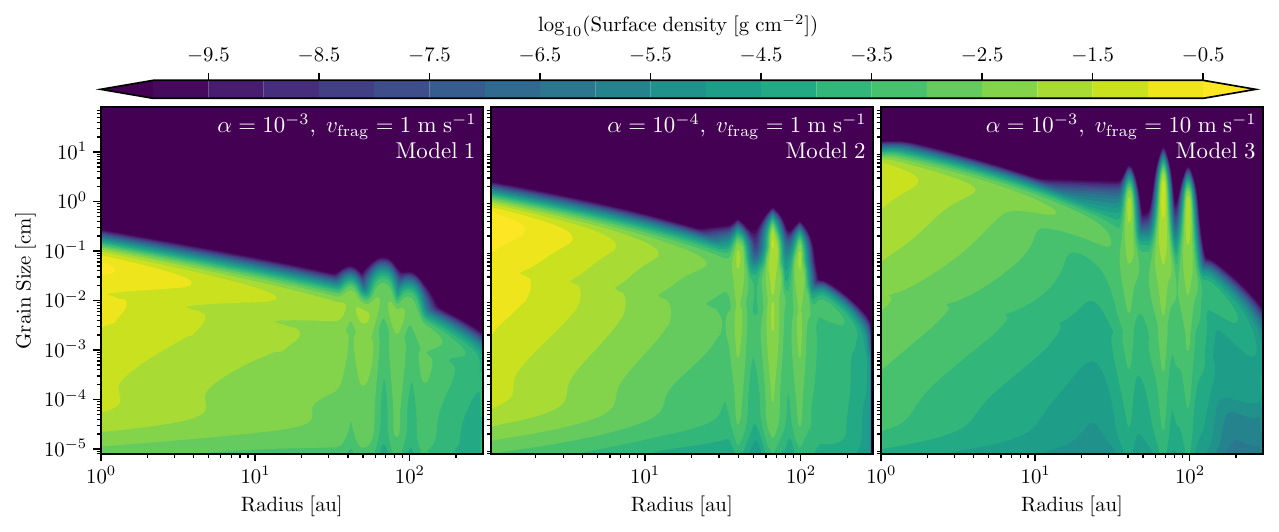}
    \caption{Dust evolution models for the disc around LkCa15. The left, middle, and right panel show the dust surface density after 1 Myrs of evolution for different velocity fragmentation and viscosity (top legend in each panel).}
    \label{fig:Evolution_models}
\end{figure*}

Thanks to the wealth of (sub-) millimetre data available for LkCa15, this target is ideal to test dust evolution models, in order to potentially constrain key parameters of disc evolution, such as disc turbulence. 

For this aim, we perform tailored models for the LkCa15 disc, using \texttt{Dustpy} version 1.0.5 \citep{Stammler_2022}. \texttt{Dustpy} simultaneously calculates the dust growth processes (dust coagulation, fragmentation, erosion) with the dynamics of the particles, including  drag (that leads to radial drift) and  diffusion. For these models, we assume a stellar mass of 1\,$M_\odot$ and we do not model the evolution of the gas, meaning that the gas remains constant during the evolution of the dust. The initial gas to dust mass ratio is 100 everywhere, and the initial grain sizes follow a Minimum Mass Solar Nebulae (MMSN) distribution with grain sizes between 0.1 and 1 micrometers. For the inclusion of the rings, we add three Gaussian rings to the unperturbed gas surface density, this means
\begin{equation} \label{eq:Gas_density_Eq}
    \Sigma^\prime_{\rm{g}}(r) = \Sigma_{\rm{g}}(r)\times[1+G_1+G_2+G_3],
\end{equation}
where
\begin{eqnarray}
G_i(r) &=& A_i\exp\left(-\frac{(r-r_{i,p})^2}{2w_i^2}\right), \\
\Sigma_{\rm{g}}(r) &=& \Sigma_0\left(\frac{r}{R_c}\right)^{-\gamma} \exp\left[-\left(\frac{r}{R_c}\right)^{2-\gamma}\right],    
\end{eqnarray}
$\gamma=1$, $R_c=160$ au, and $\Sigma_0$ is taken such that the disc mass is $0.1\,M_\odot$ \citep{Jin_2019}. The radial grid ranges from 1 to 300 au, and it is logarithmically spaced in 300 cells. The rings are centred at $r_{p,1}=42\,$au, $r_{p,2}=69\,$au, $r_{p,3}=101\,$au; as constrained from observations. The width of the Gaussian perturbation $w_i$ is the local scale height at each of the peak locations. Based on the contrast between the three rings, the amplitudes of the perturbations are assumed to be $A_1=1$, $A_2=5$, and $A_3=3$.

As rings are perturbations of the gas surface density, this translates to pressure bumps \citep{Pinilla_2012}. Inside pressure bumps, the radial drift is reduced and particles get trap. As a consequence, the maximum grain size within the rings is set by the fragmentation barrier \citep{Birnstiel_2012}, which is given by 

\begin{equation}
        a_{\rm{frag}}=\frac{2}{3\pi}\frac{\Sigma_g}{\rho_s \alpha}\frac{v_{\rm{frag}}^2}{c_s^2},
  \label{afrag}
\end{equation}

\noindent where $\rho_s$ is the volume density of the particles, which we take either to be 1.6 or 0.85 g\,cm$^{-3}$ depending on the opacities that are assumed afterwards for the radiative transfer models (DSHARP or Ricci's opacities, respectively), $\alpha$ is the disc viscosity \citep{Shakura_1973} that controls the dust diffusion, turbulence, and dust settling in the dust evolution models \citep{Pinilla_2021}, and $v_{\rm{frag}}$ is the  velocity threshold for which particles fragment (fragmentation velocity).  The last two parameters, $\alpha$ and  $v_{\rm{frag}}$ are still highly unconstrained from observations. Detailed analysis of dust distribution of a handful of highly-inclined discs suggest that the degree of dust settling is high, implying low values of  $\alpha\lesssim10^{-4}$ \citep{Villenave_2022, Rosotti_2023}. The maximum grain size obtained from the multi-wavelength analysis presented in Section \ref{sec:DustProperties} is around 1-2\,cm and 0.04-0.06\,cm within the rings using the DSHARP or Ricci opacities, respectively. Base on these constraints, we assume two values of $v_{\rm{frag}}=1, 10$\,m\,s$^{-1}$ and two values of $\alpha=10^{-4}, 10^{-3}$, which gives values of $a_{\rm{frag}}$ similar to those computed from the multi-wavelength analysis. We discard the combination of  $v_{\rm{frag}}=10$\,m\,s$^{-1}$ and $\alpha=10^{-4}$, as this leads to $a_{\rm{frag}}$ within in the rings higher than 100\,cm, contradicting the observations. Previous dust evolution models have demonstrated that in such cases, grains grow very efficient to boulders within the rings, decreasing the observed millimetre fluxes and underestimating values of typical millimetre fluxes of protoplanetary discs \citep{Pinilla_2020}.

Figure \ref{fig:Evolution_models} shows the dust density distributions after 1\,Myr of evolution, as a function of grain size, and distance from the star, for the three cases considered in this work (see model number in the top of each panel). This figure shows how different values of $v_{\rm{frag}}$ and $\alpha$ lead to different maximum grain size across the disc, and hence on different levels of dust trapping. For example, $v_{\rm{frag}}=1$\,m\,s$^{-1}$ and $\alpha=10^{-3}$ is the case where $a_{\rm{frag}}$ is the lowest, implying less trapping within the rings and more dust diffusion across the gaps. This implies less contrast in the dust density between rings and gaps. When $a_{\rm{frag}}$ is higher, the trapping is more efficient and the contrast between rings and gaps increases.

To test whether the observations favour any of the explored values of $a_{\rm{frag}}$ and $v_{\rm{frag}}$, we combine the results from dust evolution models with radiative transfer calculations to create synthetic images using \texttt{RADMC3D} \citep{Dullemond_2012}, as has been done in several previous studies \citep[e.g.,][]{Pohl_2017, Pinilla_2021, Pinilla_2024}. The dust volume density input into \texttt{RADMC3D} accounts for the degree of settling for different values of $\alpha$. In these models, the stellar temperature is assumed to be 5772\,K. For our calculations, we assume the same radial grid than the dust evolution models and a grid in the vertical direction that has 64 cells. The number of photons is set to $1 \times 10^{7}$, with an additional $5 \times 10^{6}$ scattering photons. From these simulations, we create synthetic images at the same wavelengths as the observations of LkCa15 and convolve them to match the same angular resolution (60 mas). For each of the three models, the DSHARP and Ricci's dust opacities are explored. 

Left images in Figure \ref{fig:Model_Maps_DSHARP} show the synthetic images for the DSHARP opacities, which match the observations more closely compared to Ricci's (shown in Appendix \ref{app:Ricci}). The disc within 25 au is removed in all cases to better visualize the disc morphology around the rings. The right panels compare the radial profiles of the brightness temperature and spectral index for each model with the observations. Model 2 ($\alpha = 10^{-4}$, $v_{\rm frag} = 1$ m s$^{-1}$) and 3  ($\alpha = 10^{-3}$, $v_{\rm frag} = 10$ m s$^{-1}$) seem to best match the brightness radial profiles and spectral indices, while Model 1 ($\alpha = 10^{-3}$, $v_{\rm frag} = 1$ m s$^{-1}$) is not able to reproduce the high contrasts between rings and gaps.

\begin{figure*}
    \centering
    \includegraphics[width=\linewidth]{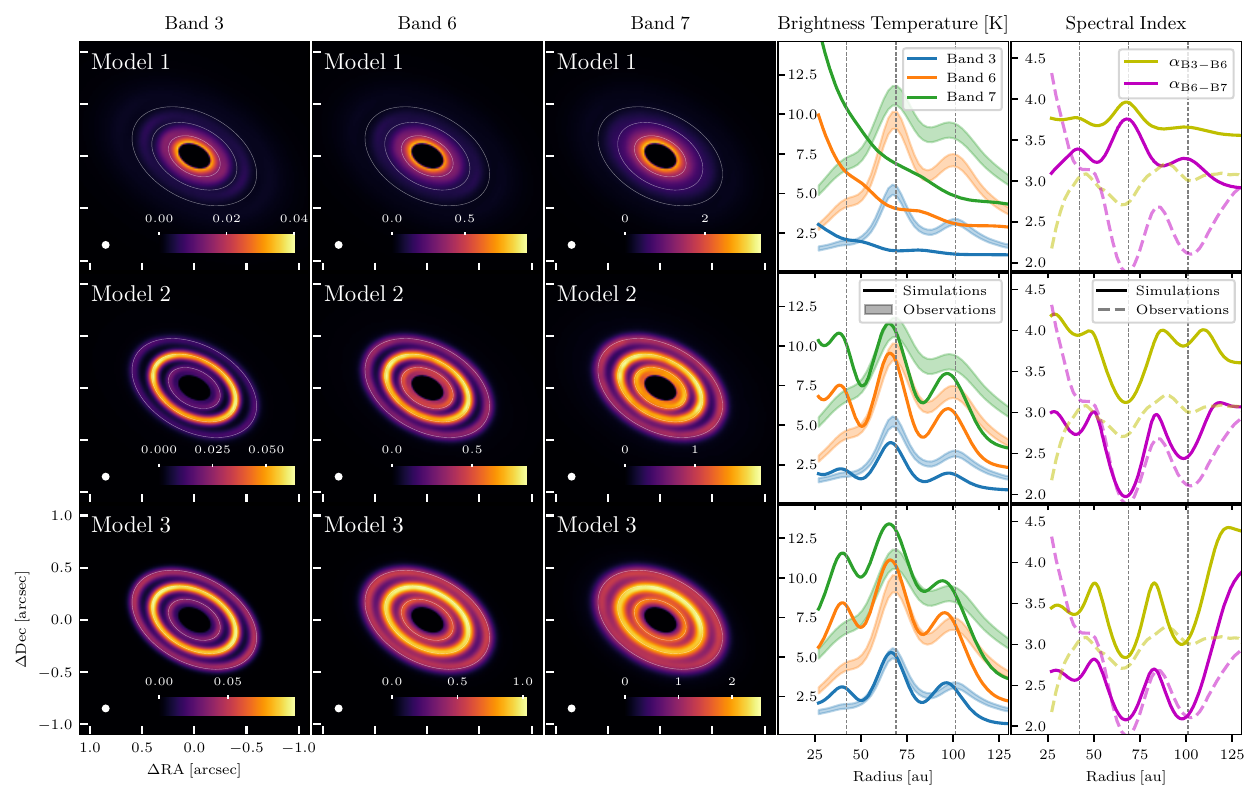}
    \caption{Simulated dust continuum emission emission for the dust evolution model 1, 2 and 3 (from top to bottom). The images on the left show the \texttt{RADMC3D} emission maps at Band 3, 6, and 7 (from left to right). The units of the colour bars in all cases are mJy beam$^{-1}$, and the ellipses show the radial position of the three bright rings at 42, 69, and 101 au for reference. The right panels show the radial profile of the brightness temperature and spectral index for each simulation (solid lines) and those from the observations (shared areas and dashed lines). The angular resolution in all cases is 60 mas, the same as the observational data in Figure \ref{fig:CleanData}.}
    \label{fig:Model_Maps_DSHARP}
\end{figure*}

As an additional test, we compute the intensity ratio between the ring at 69 au and the gaps at 51 au and 83 au (I(B69)/I(D51), I(B69)/I(D83)), and between the ring at 101 au and the gap at 83 au (I(B101)/I(D83)) for all models, and they are compared with the observations. These values are summarized in Table \ref{tab:Ring_properties}. This table also includes the fit to the FWHM of each ring between 25 au and 125 au, following the same methodology as in Figure \ref{fig:Widths}. Additionally, we include the constraints to the integrated spectral index for each model, and the total flux in Band Q and Band 9 estimated from the \texttt{RADMC3D} maps at these wavelengths. We highlight in bold the model that is closest to the observed value for each parameter, which, in most cases, corresponds to Model 3 ($\alpha = 10^{-3}$, $v_f = 10$ m s$^{-1}$).

We remark that the maximum grain size in our modelling is set by the fragmentation limit \citep{Birnstiel_2012}, which is directly proportional to the assumed gas surface density. Therefore, if the gas mass would be one order of magnitude lower \citep[as estimated by ][]{Sturm_2023}, the maximum grain size would be one order of magnitude lower, favouring models with lower turbulence and higher fragmentation velocities. Recently, as part of the exoALMA project, the gas disc mass of LkCa15 was modelled by simultaneously fitting the $^{12}$CO and $^{13}$CO rotation curves (Longarini et al. submitted), and it is found to be in agreement with the estimate in \cite{Jin_2019}, supporting our specific choice of the gas disc mass.

\begin{table*}
    \centering
    \caption{disc morphology properties of the \texttt{RADMC3D} simulations in Figure \ref{fig:Model_Maps_DSHARP}.}    
    \begin{tabular}{c|ccc|ccc|ccc|ccc}
    \hline
    \multirow{2}{*}{Parameter}  & \multicolumn{3}{c|}{Observed} & \multicolumn{3}{c|}{Model 1} & \multicolumn{3}{c|}{Model 2} & \multicolumn{3}{c}{Model 3}\\
    & B7 & B6 & B3 & B7 & B6 & B3 & B7 & B6 & B3 & B7 & B6 & B3 \\
    \hline \hline
    I(B69)/I(D51) & 2.2 & 3.7 & 4.4 & 0.6 & 0.4 & 0.4 & \textbf{2.3} & \textbf{3.4} & 6.5 & 1.6 & 2.2 & \textbf{4.5}\\
    I(B69)/I(D83) &  1.6 & 2.1 & 2.9 & \textbf{1.4} & 1.2 & \textbf{0.9} & 3.4 & 5.0 & 9.4 &  2.0 & \textbf{2.5} & 5.3\\
    I(B101)/I(D83) &  1.0 & 1.3 & 1.4 & 0.5 & 0.5 & 0.5 &  1.7 & 2.1 & 2.2 &  \textbf{1.0} & \textbf{1.2} & \textbf{2.1}\\
    FWHM$_{42}$ [au] & 24.1 $\pm$ 1.1 & 22.5 $\pm$ 0.9 & 14.6 $\pm$ 2.8 & - & - & - & 10.8 & 10.9 & 10.8 & \textbf{23.3} & \textbf{16.8} & \textbf{14.9} \\
    FWHM$_{69}$ [au] & 21.1 $\pm$ 0.2 & 19.1 $\pm$ 0.1 & 17.1 $\pm$ 0.1 & - & - & - & 16.4 & 15.5 & 14.4 & \textbf{25.3} & \textbf{18.9} & \textbf{15.5} \\    
    FWHM$_{101}$ [au]& 35.4 $\pm$ 0.5 & 29.9 $\pm$ 0.4 & 24.8 $\pm$ 0.5 & - & - & - & \textbf{23.9} & 21.5 & \textbf{19.9} & 18.1 & \textbf{23.4} & 18.6 \\  
    \hline
    $\overline{\alpha}_{\rm B3-B6}$ (25-125 au) & \multicolumn{3}{c|}{$3.0 \pm 0.9$} & \multicolumn{3}{c|}{3.7} & \multicolumn{3}{c|}{3.7} & \multicolumn{3}{c}{\textbf{3.6}}\\
    $\overline{\alpha}_{\rm B6-B7}$ (25-125 au) & \multicolumn{3}{c|}{$2.5 \pm 0.4$} & \multicolumn{3}{c|}{3.2} & \multicolumn{3}{c|}{\textbf{2.7}} & \multicolumn{3}{c}{\textbf{2.7}}\\
    \hline 
    Flux$_{\rm BQ} $ (25-125 au) [mJy] & \multicolumn{3}{c|}{0.72 $\pm$ 0.07} & \multicolumn{3}{c|}{0.08} & \multicolumn{3}{c|}{0.12} & \multicolumn{3}{c}{\textbf{0.43}} \\
    Flux$_{\rm B9}$ (25-125 au) [Jy] & \multicolumn{3}{c|}{1.22 $\pm$ 0.24} & \multicolumn{3}{c|}{\textbf{1.13}} & \multicolumn{3}{c|}{1.04} & \multicolumn{3}{c}{1.45} \\
    \hline
    \end{tabular}
    \newline
    \label{tab:Ring_properties}
    Model 1: $\alpha = 10^{-3}$, $v_{\rm frag} $ = 1 m s$^{-1}$. Model 2: $\alpha = 10^{-4}$, $v_{\rm frag} $ = 1 m s$^{-1}$. Model 3: $\alpha = 10^{-3}$, $v_{\rm frag} $ = 10 m s$^{-1}$.
    Bold text is used to highlight the model that is closest to the observed parameters. The FWHMs are not reported for Model 1 because a clear ring-like morphology is not present in the intensity profiles.
\end{table*}

%----------------------------------------------------------
\section{Discussion}\label{sec:discussion}

\label{sec:NoLagrangian}
\subsection{No Lagrangian structures in Band 3}
No azimuthal asymmetries above 3$\sigma$ were detected in the Band 3 dust continuum observations at the locations of the Lagrangian structures observed in Band 6 and 7 (Section \ref{sec:residuals}). These structures (a clump and an arc) were detected at a 10$\sigma$ significance level in both Band 6 and 7 \citep{Long_2022}. Extrapolating the clump and arc fluxes to 97.5 GHz using a spectral index of 3.0 (computed at the asymmetry locations), results in an expected intensity of $\sim 5 \ \mu$Jy beam$^{-1}$. This value is approximately equal to the rms noise of the Band 3 observations (Table \ref{tab:observations}), making it difficult to distinguish them from thermal noise.

The non-detection of structures around the Lagrangian points in Band 3 provides insights into dust trapping at the Lagrangian points. In Band 3, it is expected that observations are more sensitive to larger particles than in Band 6.
If dust trapping is happening in the Lagragian points, those large grains would be bright in Band 3, as predicted by numerical simulations \citep{Drkazkowska_2019, Montesinos_2020}. However, since this is not observed with the current sensitivity, it is likely that only smaller grains (sub-millimetre-sized), which are bright in
ALMA Bands 6 and 7 \citep{Draine_2006}, are dominating the emission of the potential Lagrangian points. A possible explanation for the absence of large grain sizes at the Lagrangian structures could be the filtering of the large dust grains by the dust trap at 69 au or the influence of the potential dust trap at 42 au, which might be stopping the reservoir of large grains moving to the Lagrangian points.

%----------------------------------------------------------
\label{sec:Efficiency}
\subsection{Dust trap efficiency}
The efficiency of dust trapping in ringed protoplanetary discs has been studied by  \cite{Rosotti_2020} and \cite{Dullemond_2018}. In the former, gas dynamics information is needed to constrain dust parameters as the Stokes parameter or the $\alpha$ turbulent parameter. In the latter, only dust continuum emission, as that in this work, is needed to study the dust trap mechanism.

As mentioned in \cite{Dullemond_2018}, the width of the radial pressure bump ($w_{p}$) originating the dust traps cannot be smaller than the width of the dust ring ($w_{\rm d}$) or the local thermal pressure scale height ($h_p$), i.e., $w_{p, \rm min} = {\rm max}(w_{\rm d}, h_p)$. For this reason, we use the Band 3 radial profiles inferred from the \textsc{Frankenstein} fit as reference, as they exhibit the narrowest dust continuum distribution among the observations in this work (Figure \ref{fig:Widths}). The widths $w_{\rm d}$ for each ring are the FWHM in Table \ref{tab:gaussians}, and they are deconvolved using their angular resolution (5.5 au) as follows:  $w_d = \sqrt{{\rm FWHM}^2 - (5.5 {\rm au})^2 } /(2 \sqrt{2 \ln(2)})$. The factor in the denominator consistently converts the FWHM to the standard deviation, as originally derived in \cite{Dullemond_2018}. On the other hand, the width of the radial pressure bump cannot exceed the radial separation of the rings \citep[as defined in ][]{Dullemond_2018}. The local pressure scale height is estimated from the temperature of a passively irradiated disc and assuming hydrostatic equilibrium \citep[Equations 5 and 6 in ][]{Dullemond_2018}.

Constraints on the gas surface density at the location of the rings are also necessary to study the dust trap mechanism. Unfortunately, a constant gas-to-dust ratio cannot be used, as the dust trapping mechanism is expected to concentrate dust grains in the rings and decrease the local gas-to-dust ratio. Low gas-to-dust ratios in the rings of LkCa15 have been inferred by \cite{Sturm_2023} using observations of several CO isotopologues and continuum data. Additionally, \cite{Sturm_2023} found gas mass estimates that are one order of magnitude lower than the results in \cite{Jin_2019}.
Therefore, different gas mass estimates may influence the conclusions regarding the dust trap mechanism.

Fortunately, \cite{Dullemond_2018} was able to study the dust trap mechanism by estimating upper and lower limits to the gas surface density using only physical constraints, independently of a specific gas surface density measurement.
The upper limit for the gas surface density is given requiring gravitationally stability (Toomre parameter $>$ 2), while the lower limit is given by the local dust surface density (i.e. it requires a gas-to-dust ratio larger than one). The results in both \cite{Jin_2019} and \cite{Sturm_2023} fall within these limits. The dust surface density constraints are taken from those computed in Section \ref{sec:DustProperties}.

These constraints are summarized in Table \ref{tab:trapping}. where we also include the constraints to the maximum grain size in Figure \ref{fig:DustProp}, and to the minimum and maximum Stokes number in the Epstein regime \citep{Epstein_1924}. The minimum and maximum value for the latter is estimated using the maximum and minimum gas surface density, respectively. The bulk densities for the DSHARP and Ricci's dust grain compositions are 1.6 and 0.85 g\,cm$^{-3}$, respectively.

The local thermal scale heights (estimated from the temperature of a passively irradiated disc) at B42, B69 and B101 are $h_p = $ 2.2, 4.0, 6.5 au, respectively, while the dust widths at Band 3 are $w_d$ = 3.5, 3.6, 6.3 au, respectively. Except for B42, the width of the dust rings are narrower than the scale heights, which can be interpreted as evidence that dust trapping must have taken place in B69 and B101 \citep{Dullemond_2018}. The B42 ring does not show a peak in dust surface density or maximum grain size either (Figure \ref{fig:DustProp}), leaving its origin still unclear.

The upper and lower limit of the ratio $\alpha_{\rm turb} / \rm St$ can now be estimated using the Equation 21 in \cite{Dullemond_2018} and the lower and upper limit for $w_p$, with a Schmidt number = 1 \citep{Johansen_2005}. Finally, lower limits for $\alpha_{\rm turb}$ can also be estimated using the lower limit on the the Stokes number. All the parameters are reported in Table \ref{tab:trapping}. The lower limits for $\alpha_{\rm turb}$ are within the typical values found in \cite{Pinte_2016, Flaherty_2018, Dullemond_2018}, which lies around $\alpha_{\rm turb} \sim 10^{-4}, 10^{-3}$.

Note that the possible Stokes number values at B101 tend to be higher than those at B69, which mainly occurs due to the lower gas surface densities in the outer ring. However, although higher Stokes numbers are allowed in the outer ring, both the dust surface density and maximum grain size inferred from the multi-wavelength analysis (Figure \ref{fig:DustProp}) in B69 are higher than those at B101. Furthermore, similar results are obtained from the three dust evolution models, as shown in Figure \ref{fig:Evolution_models}. 

All the observational and model constraints indicate that higher dust surface densities and grain sizes are present in B69 compared with those in B101. The fact that the slope of the FWHM of the B101 ring decreases faster than in B69 (Figure \ref{fig:Widths}) can be associated to an optically depth effect. We remark that interpreting a higher slope as evidence of higher dust trapping efficiency is only valid in the optically thin regime, where the dust rings directly trace the mass distribution of solids. Optical depth effects may contribute to broadening the observed dust ring width, as in the case of the B69 ring (Figure \ref{fig:OptDepths}).

The radial position of the B101 ring differs slightly between Bands 7, 6, and 3 (Figures \ref{fig:CleanData}, \ref{fig:VisData}). Although the peak position appears to increase with wavelength, the difference between the radial position of B101 at Bands 7 and 3 is only about 3 au (Table \ref{tab:gaussians}). This difference is smaller than the angular resolution and limits robust conclusions about the origin of the apparent shift, which could be attributed to radial drift and/or optical depth effects. 

The hydrodynamical simulations by \cite{Facchini_2020} suggest that the rings in LkCa15 can be attributed to the presence of an embedded planet, though their specific appearance and the number of rings are highly sensitive to the disc's thermodynamics. This potentially can also explain the tentative faint ring observed beyond 1 arcsec (157.2 au) in Band 3, 6, 7 (Figures \ref{fig:CleanData}, \ref{fig:VisData}), and Band Q \citep{Sierra_2024b}.

\begin{table}
    \centering
    \caption{Constraints on dust traps in the rings of LkCa15.}
    \label{tab:trapping}    
    \begin{tabular}{c|ccc|ccc}
    \hline 
    \multirow{2}{*}{Parameter} & \multicolumn{3}{c|}{DSHARP Opacities} & \multicolumn{3}{c}{Ricci's Opacities} \\
      & B42 & B69 & B101 & B42 & B69 & B101 \\
     \hline \hline 
     $w_{p, \rm min}$ [au] & 3.5 & 4.0 & 6.5 & 3.5 & 4.0 & 6.5\\
     $w_{p, \rm max}$ [au] & 18  & 36  & 36 & 18  & 36  & 36 \\
     $\Sigma_{\rm{g, min}}$ [g cm$^{-2}$] & 3.0E-2 & 6.8E-1 & 1.7E-1 & 1.7E-3 & 1.9E-2 & 8.2E-3\\
     $\Sigma_{\mathrm{g, max}}$ [g cm$^{-2}$] & 5.4E1 & 2.3E1 & 1.2E1 & 5.4E1 & 2.3E1 & 1.2E1 \\
     $a_{\mathrm{max}}$ [cm] & 4.5E-1 & 5.9E-1 & 8.3E-1 & 3.7E-2 & 5.8E-2 & 4.0E-2\\
     $\rm St_{{min}}$ & 2.2E-2 & 6.7E-2 & 1.8E-1 & 9.2E-4 & 3.4E-3 & 4.4E-3\\ 
     $\rm St_{{max}}$ & 3.9E1 & 2.3E0 & 1.3E1 & 2.9E1 & 4.1E0 & 6.5E0\\
     $(\alpha_{\rm turb}/\rm St)_{\rm min}$ & 3.9E-2 & 1.0E-2 & 3.2E-2 & 1.1E-1 & 2.9E-2 & 9.4E-2 \\     
     %(\alpha_{\rm turb}/\rm St)_{\rm max}$ & - & 4.3 & 1.6E1 & - & 4.3 & 1.6E1 \\
     $\alpha_{\rm turb, min}$ & 8.6E-4 & 6.7E-4  & 5.7E-3 & 1.0E-4 & 9.9E-5 & 4.1E-4 \\ 
     %$\alpha_{\rm turb, max}$ & -      & 8.2E0   & 1.6E2  & - &  3.4E1 & 2.4E2 \\
    \hline
    \end{tabular}
\end{table}

%--------------------------------------------------------------------
\subsection{Dust opacities}\label{sec:opacities}

Dust opacity significantly influences the derived properties of protoplanetary discs. Variations in assumptions about their composition \citep[e.g.,][]{Birnstiel_2018}, grain sizes \citep[e.g.,][]{Draine_2006}, porosity \citep[e.g.,][]{Kirchschlager_2019, Zhang_2023, Liu_2024}, shape \citep[e.g.,][]{Tazaki_2018, Kirchschlager_2020}, among other factors, can result in absorption and scattering opacities that differ by orders of magnitude, as well as varied spectral properties, which directly impact the estimation of dust mass.

In this work we tested the DSHARP and Ricci's opacities (Figure \ref{fig:Opacities}), two of the most widely used dust opacities in protoplanetary disc studies. Although both compositions contain water ices and astronomical silicates, the inclusion of carbon grains in Ricci’s opacities (instead of refractory organics in DSHARP) increases absorption opacity by an order of magnitude at millimetre wavelengths and reduces the albedo. This leads to less massive discs when inferring dust surface density from multi-wavelength observations (Figure \ref{fig:DustProp}), or increases total intensity for a fixed dust surface density in simulations (Figure \ref{fig:Model_Maps_DSHARP} and \ref{fig:Model_Maps_RICCI}).

The gas-to-dust ratio for the DSHARP and Ricci's opacity models is approximately 132 and 2500, respectively, \citep[assuming a gas mass of 0.1 M$_{\odot}$,][]{Jin_2019}. This ratio is particularly high for Ricci's opacities, while it falls within the typical expected values for the Interstellar Medium (ISM) and protoplanetary discs \citep{Ansdell_2016, Zhang_2021} for the DSHARP opacities. 
On the other hand, the small grain sizes and dust surface densities derived from Ricci's opacities have significant implications for the expected emission at very long wavelengths. Specifically, the flux of LkCa15 at VLA Band Q (44 GHz)\footnote{The disc was observed as part of the Disks@EVLA Collaboration, \href{https://safe.nrao.edu/evla/disks}{https://safe.nrao.edu/evla/disks}} between 25 au and 125 au is approximately 0.72 mJy.
The expected flux at Band Q (within the same radial extent) based on the inferred dust properties shown in Figure \ref{fig:DustProp} is $\sim$ 0.70 mJy and 0.13 mJy for the DSHARP and Ricci's opacity models, respectively. The latter value is particularly low compared to the observed flux, likely due to a combination of low dust surface densities and the reduced opacity of sub-mm grain sizes at $\lambda = 7$ mm \citep{Draine_2006}.

\begin{figure}
    \centering
    \includegraphics[width=\linewidth]{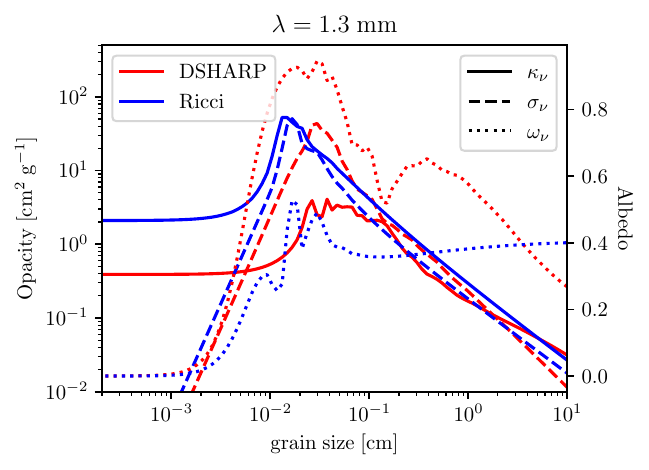}
    \caption{Dust opacity and albedo at $\lambda = 1.3$ mm for the DSHARP (red) and Ricci (blue) composition. The solid, dashed, and dotted lines show the absorption, scattering, and albedo coefficients, respectively.}
    \label{fig:Opacities}
\end{figure}

%----------------------------------------------------------
\section{Conclusions}\label{sec:conclusions}
The new Band 3 dust continuum observations ($\lambda = 3$ mm) of the disc around LkCa15 offer unprecedented insights into the distribution of solids within its ring structures. The Band 3 data were observed at the highest angular resolution of ALMA's capabilities, achieving a resolution of 60 mas (9.4 au) when imaging with CASA, or 35 mas (5.5 au) when modelling the visibilities.

The bright rings B42, B69, and B101, previously resolved at shorter wavelengths, are confirmed, revealing deeper gaps and narrower ring structures compared to the dust continuum observations at Band 6 ($\lambda = 1.3$ mm) and Band 7 ($\lambda = 0.88$ mm). The width of the B42, B69, and B101 rings increases with frequency at rates of 0.31, 0.22, and 0.41 mas GHz$^{-1}$, respectively, consistent with signatures predicted by dust trap models.
Additionally, the Band 3 data reveals a bright compact emission in the inner disc, which given its low spectral indices between 1.3 and 3 mm, and 3 mm and 3 cm ($\alpha_{\rm 1.3-3mm} = 0.30 \pm 0.37$, $\alpha_{\rm 3mm-3m} = [-0.1, 0.0]$), may be tracing a combination of dust and free-free emission. The origin of the latter is consistent with an ionised jet, as reported by \cite{Rota_2024} in the inner discs of 10 transitional discs with similar spectral index properties.

Although most of the dust continuum data at Band 3 looks axisymmetric, there are some azimuthal residuals around the $\sim 3 \sigma$ level. One of these residuals is detected with an SNR of 3.5 in the 42 au orbit, near the boundary of the Northern arc identified in Bands 7 and 6 by \cite{Long_2022}. However, no Band 3 azimuthal asymmetries are detected at the Lagrangian structures observed in Bands 7 and 6. On the other hand, two point sources were also identified with an SNR of 2.5 in the orbits at 16.5 and 20.9 au, where two possible planets were proposed based on H$\alpha$ observations in \cite{Sallum_2015}, although these were later explained by scattered light \citep{Sallum_2023}. The position of the outer point source at 20.9 au aligns with the expected orbital motion of these structures around a 1.2 M$_{\odot}$ star.

The resolved multi-wavelength modelling of the dust continuum emission at Bands 3, 6, and 7 is used to constrain the radial profiles of dust temperature, dust surface density, and maximum grain size. The B69 and B101 rings show peak values for dust surface density and maximum grain size when using DSHARP or Ricci's opacities, confirming the signatures predicted by dust trapping models. Although the width of the B42 ring increases with frequency, there is not a robust evidence of a dust trap when modelling the multi-wavelength data, and its origin remains unclear. These dust trap signatures at B69 and B101 are confirmed using the analytical prescription tests in \cite{Dullemond_2018}.

The dust surface density and maximum grain size computed using DSHARP opacities are an order of magnitude larger than those obtained from Ricci's opacities. This directly impacts the total dust mass, which is estimated at $7.6 \times 10^{-4}$ M$_{\odot}$ for DSHARP opacities and $0.3 \times 10^{-4}$ M$_{\odot}$ for Ricci's opacities. These dust masses are factors of 132 and 2500 smaller than previous estimates of the total gas mass \citep[0.1 M$_{\odot}$, ][]{Jin_2019}, suggesting that properties inferred from the DSHARP opacities are more consistent with previous constraints on the gas-to-dust mass ratio in several protoplanetary discs \citep{Ansdell_2016, Zhang_2021}. 

The dust evolution models of LkCa15 in this work show that a 0.1 M$_{\odot}$ gaseous disc with Gaussian rings at B42, B69, and B101, a viscosity $\alpha = 10^{-3}$, a fragmentation velocity of $v_f = 10$ m s$^{-1}$, and DSHARP opacities, is the best model to reproduce the contrast between rings and gaps of the intensity radial profiles at Bands 3, 6, and 7, as well as the integrated fluxes at Bands Q and 9, and the spectral indices. In both the dust evolution models and the dust properties inferred from observations, the efficiency of dust trapping in the B69 ring appears to be the highest, due to the larger dust mass and greater maximum grain sizes reached in this ring compared to B101.

After computing the dust continuum maps of the dust evolution models using the DSHARP and Ricci's opacities, we found that the former do a better work in reproducing the observations, also suggesting that the DSHARP composition works better when modelling the millimetre dust continuum emission of LkCa15.

%----------------------------------------------------------
\section*{Acknowledgments}
We are very thankful for the thoughtful suggestions of the anonymous referee that helped to improve our manuscript significantly.
A.S. and P.P. acknowledge funding from the UK Research and Innovation (UKRI) under the UK government’s Horizon Europe funding guarantee from ERC (under grant agreement No 101076489).
MB has received funding from the European Research Council (ERC) under the European Union's Horizon 2020 research and innovation programme (PROTOPLANETS, grant agreement No. 101002188).
PC acknowledges support by the ANID BASAL project FB210003.
Support for F.L. was provided by NASA through the NASA Hubble Fellowship grant \#HST-HF2-51512.001-A awarded by the Space Telescope Science Institute, which is operated by the Association of Universities for Research in Astronomy, Incorporated, under NASA contract NAS5-26555.  
C.C.-G. acknowledges support from UNAM DGAPA-PAPIIT grant IG101224 and from CONAHCyT Ciencia de Frontera project ID 86372.
This paper makes use of the following ALMA data: 2012.1.00870.S, 2015.1.00118.S, 2018.1.00945.S, 2018.1.01255.S, 2018.1.00350.S, and 2022.1.01216.S. ALMA is a partnership of ESO (representing its member states), NSF (USA) and NINS (Japan), together with NRC (Canada), MOST and ASIAA (Taiwan), and KASI (Republic of Korea), in cooperation with the Republic of Chile. The Joint ALMA Observatory is operated by ESO, AUI/NRAO and NAOJ.

%----------------------------------------------------------
\section*{Software}
This work made use of the following software:
Astropy \citep{astropy:2013, astropy:2018},
CASA \citep{McMullin_2007}, Dustpy \citep{Stammler_2022}, Emcee \citep{Foreman_2013}, Frankenstein \citep{Jennings_2020}, Gofish \citep{GoFish}, Matplotlib \citep{Matplotlib_2007}, Numpy \citep{Numpy_2020}, RADMC3D \citep{Dullemond_2012}, Scipy \citep{SciPy_2020}.

%----------------------------------------------------------
\section*{Data availability}
The self-calibrated data underlying this article will be shared on reasonable request to the corresponding author. The non-calibrated data is publicly available at \href{https://almascience.nrao.edu/aq/}{https://almascience.nrao.edu/aq/} using the project code 2022.1.01216.S.

%%%%%%%%%%%%%%%%%%%% REFERENCES %%%%%%%%%%%%%%%%%%

% The best way to enter references is to use BibTeX:

\bibliographystyle{mnras}
\bibliography{main} % if your bibtex file is called example.bib
%\input{main.bbl}

%%%%%%%%%%%%%%%%%%%%%%%%%%%%%%%%%%%%%%%%%%%%%%%%%%

%%%%%%%%%%%%%%%%% APPENDICES %%%%%%%%%%%%%%%%%%%%%

\appendix
%----------------------------------------------------------
\section{disc centring} \label{app:Multi_Residuals}

Errors in the disc geometry and offset with respect to the Phase Center can create non-axi-symmetric structures in the residual maps of the dust continuum modelling \citep{Andrews_2021}. As mentioned in Section \ref{sec:results}, we estimated the offset of the disc by fitting the iso-contour ellipse that follows the main ring B69 and obtain $\Delta \rm RA = 15.7$ mas,  $\Delta \rm Dec = -6.9$ mas. On other side, the offset was also estimated by minimizing the imaginary part of the visibilities \citep{Isella_2019}, obtaining $\Delta \rm RA = 15.3$ mas, $\Delta \rm Dec = 12.9$ mas. All these values are on the order of one pixel size. A \textsc{Frankenstein} model was calculated in both cases, and the residual maps were computed using the same imaging parameters as in Figure \ref{fig:Robust-Residuals}. The residual maps are shown in Figure \ref{fig:Center-Residual}, where it is clear that the residual structures are minimized at the disc phase centre. Therefore, we assume $\Delta \rm RA = \Delta \rm Dec = 0$ when fitting the Band 3 data.

\begin{figure*}
    \centering
    \includegraphics[width=\linewidth]{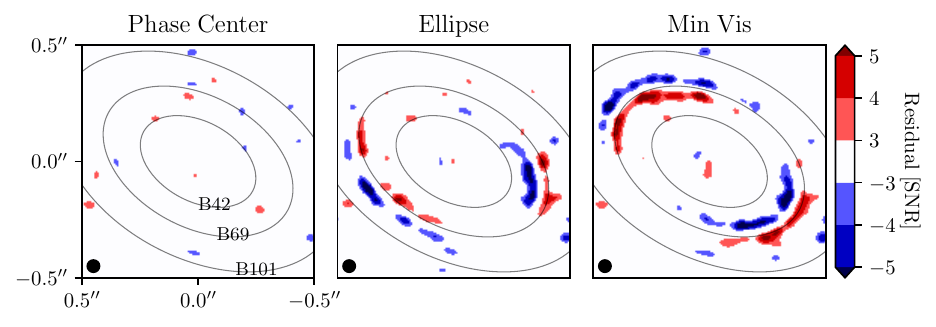}
    \caption{Residual map of the Band 3 dust continuum observations of LkCa15 after subtracting an axi-symmetric model centred at the Phase Center (left), in the centre computed from an ellipse fit (middle), and the centre computed from minimizing the imaginary part of the visibilities (right). The colour bar is the same for all panels. The ellipses show the radial position of the three bright rings at 42, 69, and 101 au. The beam size is shown in the bottom left corner of each panel.}
    \label{fig:Center-Residual}
\end{figure*}

%----------------------------------------------------------
\section{Radiative transfer models for Ricci's opacities}\label{app:Ricci}

The radiative transfer models computed with \texttt{RADMC3D} using the Ricci's opacities for the three dust evolution models in Figure \ref{fig:Evolution_models} are shown in Figure \ref{fig:Model_Maps_RICCI}. The brightness temperatures for all bands are significantly higher compared to those from the DSHARP opacities (Figure \ref{fig:Model_Maps_DSHARP}) and the observations. The main difference between Ricci's and DSHARP opacities comes from their composition, as the carbonaceous grains in the former have a higher absorption opacity than the refractory organic grains in the DSHARP mixture \citep{Birnstiel_2018}.

\begin{figure*}
    \centering
    \includegraphics[width=\linewidth]{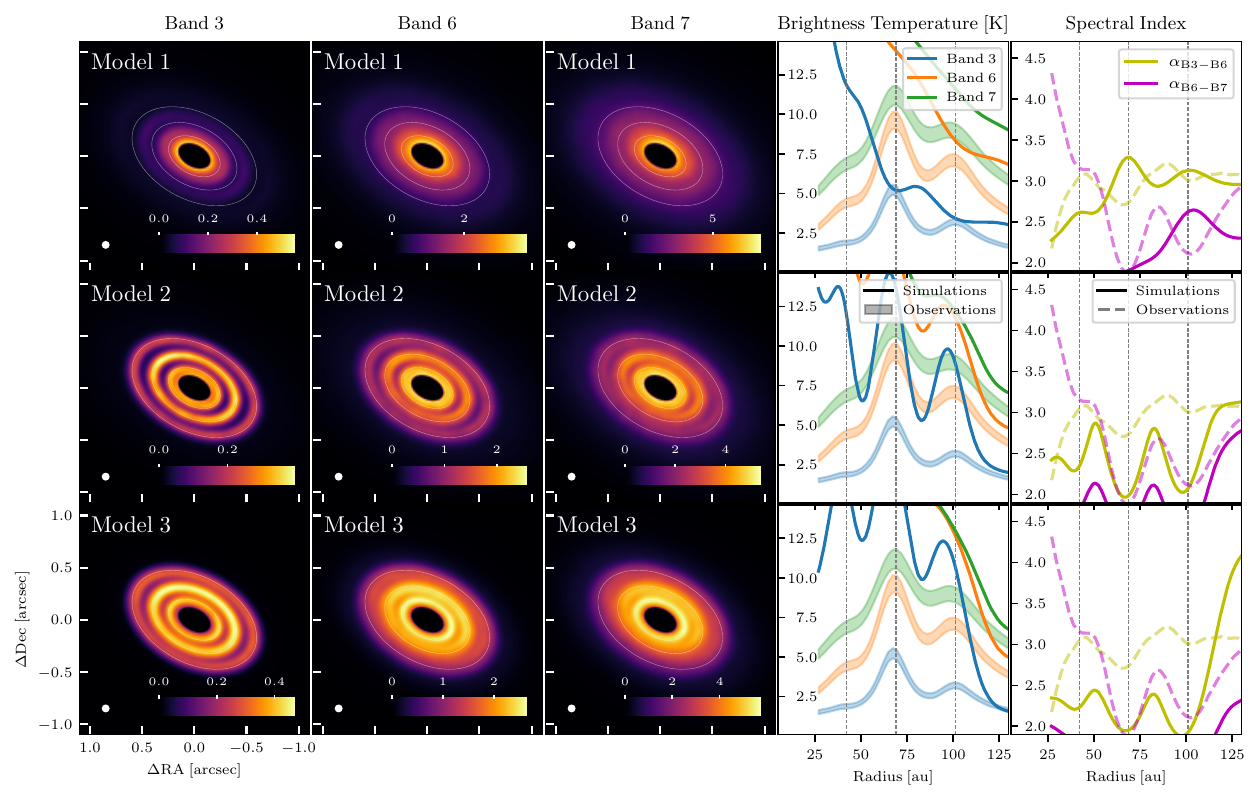} 
    \caption{Same as Figure \ref{fig:Model_Maps_DSHARP}, but using the Ricci's opacities.}
    \label{fig:Model_Maps_RICCI}
\end{figure*}

%%%%%%%%%%%%%%%%%%%%%%%%%%%%%%%%%%%%%%%%%%%%%%%%%%

% Don't change these lines
\bsp	% typesetting comment
\label{lastpage}
\end{document}